\documentclass[twocolumn]{aastex63}

\newcommand{\mysc}[1]{\textrm{\textsc{#1}}}

\usepackage{xcolor}

\received{2020 November 15}
\revised{2021 January 16}
\accepted{2021 January 21}
\submitjournal{ApJ}

\shorttitle{The VLA Frontier Fields Survey}
\shortauthors{Heywood et al.}

\begin{document}

\title{The VLA Frontier Fields Survey:\\ Deep, High-resolution Radio Imaging of the MACS Lensing Clusters at 3 and 6 GHz}

\correspondingauthor{Ian Heywood}
\email{ian.heywood@physics.ox.ac.uk}

\author[0000-0002-0786-7307]{I.~Heywood}
\affil{Astrophysics, Department of Physics, University of Oxford, Keble Road, Oxford, OX1 3RH, UK}
\affil{Department of Physics and Electronics, Rhodes University, PO Box 94, Makhanda, 6140, South Africa}
\affil{South African Radio Astronomy Observatory, 2 Fir Street, Observatory, 7925, South Africa}

\author{E.~J.~Murphy}
\affil{National Radio Astronomy Observatory, 520 Edgemont Road, Charlottesville, VA 22903, USA}

\author{E.~F.~Jim\'{e}nez-Andrade}
\affil{National Radio Astronomy Observatory, 520 Edgemont Road, Charlottesville, VA 22903, USA}

\author{L.~Armus}
\affil{Infrared Processing and Analysis Center, MC 314-6, 1200 E. California Boulevard, Pasadena, CA 91125, USA}

\author{W.~D.~Cotton}
\affil{National Radio Astronomy Observatory, 520 Edgemont Road, Charlottesville, VA 22903, USA}
\affil{South African Radio Astronomy Observatory, 2 Fir Street, Observatory, 7925, South Africa}

\author{C.~DeCoursey}
\affil{Department of Astronomy, Bryant Space Science Center, 1772 Stadium Road, Gainesville, FL 32611, USA}
\affil{National Radio Astronomy Observatory, 520 Edgemont Road, Charlottesville, VA 22903, USA}

\author{M.~Dickinson}
\affil{National Optical Astronomy Observatories, 950 N Cherry Avenue, Tucson, AZ 85719, USA}

\author{T.~J.~W.~Lazio}
\affil{Jet Propulsion Laboratory, California Institute of Technology, 4800 Oak Grove Drive, Pasadena, CA 91109, USA}

\author{E.~Momjian}
\affil{National Radio Astronomy Observatory, P.O. Box O, Socorro, NM 87801, USA}

\author{K.~Penner}
\affil{National Radio Astronomy Observatory, 520 Edgemont Road, Charlottesville, VA 22903, USA}

\author{I. Smail}
\affil{Centre for Extragalactic Astronomy, Department of Physics, Durham University, South Road, Durham DH1 3LE, UK}

\author{O.~M.~Smirnov}
\affil{Department of Physics and Electronics, Rhodes University, PO Box 94, Makhanda, 6140, South Africa}
\affil{South African Radio Astronomy Observatory, 2 Fir Street, Observatory, 7925, South Africa}



\begin{abstract}

The Frontier Fields project is an observational campaign targeting six galaxy clusters, with the intention of using the magnification provided by gravitational lensing to study galaxies that are extremely faint or distant. We used the \emph{Karl G. Jansky} Very Large Array (VLA) at 3 and 6 GHz to observe three Frontier Fields: MACSJ0416.1$-$2403 ($z$~=~0.396), MACSJ0717.5+3745 ($z$~=~0.545), and MACSJ1149.5+2223 ($z$~=~0.543). The images reach noise levels of $\sim$1 $\mu$Jy beam$^{-1}$ with sub-arcsecond resolution ($\sim$2.5 kpc at $z$~=~3), providing a high-resolution view of high-$z$ star-forming galaxies that is unbiased by dust obscuration. We generate dual-frequency continuum images at two different resolutions per band, per cluster, and derive catalogs totalling 1966 compact radio sources. Components within the areas of \emph{Hubble} Space Telescope and Subaru observations are cross-matched, providing host galaxy identifications for 1296 of them. We detect 13 moderately-lensed (2.1~$<$~$\mu$~$<$ 6.5) sources, one of which has a demagnified peak brightness of 0.9 $\mu$Jy beam$^{-1}$, making it a candidate for the faintest radio source ever detected. There are 66 radio sources exhibiting complex morphologies, and 58 of these have host galaxy identifications. We reveal that MACSJ1149.5+2223 is not a cluster with a double relic, as the western candidate relic is resolved as a double-lobed radio galaxy associated with a foreground elliptical at $z$~=~0.24. The VLA Frontier Fields project is a public legacy survey. The image and catalog products from this work are freely available.

\end{abstract}




\section{Introduction} 
\label{sec:intro}

The Frontier Fields project was conceived around the use of the \emph{Hubble} Space Telescope (\emph{HST}) and \emph{Spitzer} space telescope to undertake a deep field imaging program, targeting six strong lensing galaxy clusters, and six parallel blank fields \citep{lotz17}. The observations take advantage of the magnification boost provided by the foreground lensing clusters, allowing us to detect galaxies that are intrinsically very faint or at redshifts that approach the cosmic reionisation epoch ($z\sim$~6 and above). The goal of the Frontier Fields project is to gather a large sample of such sources in order to further understand star formation processes in the early Universe, via measurements of the stellar mass, star-formation rates, and structure of distant galaxies.

Other ground and space-based facilities have also targeted these fields as part of the Frontier Fields campaign, and in this paper we present the data release from a survey with the \emph{Karl G.~Jansky} Very Large Array (VLA). Radio observations have a singular advantage in the study of extragalactic star formation processes, in that they are not susceptible to the dust obscuration that biases observational probes of star formation at ultraviolet, optical, and infrared wavelengths. Furthermore, radio is an excellent way to probe active galactic nuclei (AGN), due to the radio emission that can arise from the AGN core itself, as well as detecting any large-scale radio jets emanating from the region around the central supermassive black hole \citep{delvecchio17}. As a way to probe the obscured star formation that is present in the Frontier Field galaxies, these radio observations will form a crucial addition in pursuit of the goals of the program. This is particularly important for studies of star formation at high redshift ($z\gtrsim$~3), as the majority fraction of star formation in massive galaxies at such epochs takes place within dusty starburst systems, but the radio luminosities of these sources lie below the typical blank-field flux density limit of current radio facilities \citep[e.g.][]{murphy11,magnelli13,wang19,dudzeviciute20}.

Our observing program has targeted the three Massive Cluster Survey \citep{ebeling01} Frontier Field clusters, namely MACSJ0416.1$-$2403 ($z$~=~0.396), MACSJ0717.5+3745 ($z$~=~0.545), and MACSJ1149.5+2223 ($z$~=~0.543; hereafter MACSJ0416, MACSJ0717 and MACSJ1149 respectively). These clusters were selected based on their favourable declinations for observations with both the VLA and ALMA, as well as (at the time the observations were proposed) the availability of ancillary data. The three target clusters are all known to host diffuse radio emission associated with processes in the intracluster medium. MACSJ0416 hosts a central radio halo \citep[e.g.][]{ogrean15}, MACSJ0717 has a very powerful central halo and a peripheral relic \citep[e.g.][]{vanweeren09,rajpurohit20}, and MACSJ1149 hosts a halo as well as a candidate double relic structure \citep[e.g.][]{bonafede12}.

We targered the three clusters using the S (2--4 GHz) and C (5--7 GHz) band receivers of the VLA, with central frequencies of 3 and 6~GHz respectively. Ninety percent of the observing was done using the most extended A-configuration of the VLA to achieve the highest possible angular resolution. Coupled with long integration times, these observational parameters enable us to potentially detect and resolve $L^{*}$ galaxies at $z\sim$~3, and to discover lensed sub-$L^{*}$ galaxies at similar redshifts and above.

This paper describes the observations, data reduction, and the production and validation of the radio data products. We have also cross-matched the radio detections with the optical data from the \emph{HST} \citep{castellano16,dicriscienzo17,shipley18} and Subaru \citep{medezinski13,umetsu14} to produce catalogs of radio components with host identifications and properties derived from the optical / near-infrared data, for each cluster. 

An overview of the radio observations and the data calibration and imaging methods is given in Section \ref{sec:obs}. The image products and the construction of the catalogs derived from them is described in Section \ref{sec:results}, along with a description of the optical cross-matching procedure and the estimation of gravitational lensing magnifications for radio sources that have a redshift measurement. Some results are discussed in Section \ref{sec:discussion}, and Section \ref{sec:conclusions} concludes the paper and provides links to the publicly available data products. The assumed cosmological model throughout this paper is $\Lambda$-CDM with H$_{0}$~=~70~km~s$^{-1}$~Mpc$^{-1}$, $\Omega_{M}$~=~0.3 and $\Omega_{\Lambda}$~=~0.7.

\section{Observations, calibration, and imaging}
\label{sec:obs}

The data\footnote{Project codes: 14A-012, 15A-282, and archival data from SF0858} were taken with the VLA in its A (77.91 hours) and C (8.34 hours) configurations using the S-band (2--4 GHz) and C-band (5--7 GHz) receivers, for a total of 86.25 hours. A summary of the radio observations is provided in Table \ref{tab:obs}. The data were taken in full polarisation mode, however, this article only presents the total intensity (Stokes I) radio images and derived products.

\begin{table*}
\begin{minipage}{175mm}
\centering
\caption{Coordinates and calibrators for each of the three target clusters, as well as the on-source integration times for each of the configuration/band pairings. Note that the A-configuration observations in S-band for MACSJ0717 include archival data previously published by \citet{vanweeren16}.}
\begin{tabular}{llccccc} \hline \hline
                 &           & MACSJ0416 & MACSJ0717 & MACSJ1149 & Band & Configuration\\ \hline
Right Ascension  & J2000     & 04h16m08.9s      & 07h17m34.0s     & 11h49m36.3s      & -& -\\
Declination      & J2000     & -24d04m28.7s     & 37d44m49.0s     & 22d23m58.1s      & -& -\\
Redshift         &           & 0.396            & 0.545           & 0.543            & -& -\\ 
Primary calibrator      &           & 3C48 & 3C147 & 3C286 & -& -\\
Secondary calibrator    &           & J0416-1851 & J0714+3534 & J1158+2450 & & \\
Integration time & [h]         & 18.08            & 15.30           & 8.35             & S & A\\
Integration time & [h]         & 1.39             & 1.39            & 1.39             & S & C\\
Integration time & [h]         & 22.26            & 6.96            & 6.96             & C & A \\
Integration time & [h]         & 1.39             & 1.39            & 1.39             & C & C\\ \hline
\end{tabular}
\label{tab:obs}
\end{minipage}
\end{table*}

Each Scheduling Block (SB) was initially processed using the NRAO VLA pipeline\footnote{{\tt \tiny \href{https://science.nrao.edu/facilities/vla/data-processing/pipeline}{https://science.nrao.edu/facilities/vla/data-processing/pipeline}}}. This is a set of scripts for \mysc{CASA} \citep[Common Astronomy Software Applications;][]{mcmullin07} package designed to perform basic calibration steps on continuum data. The pipeline performs flagging of data due to antenna shadowing, visibility amplitudes that are exactly zero, and the initial integrations following antenna slewing. A first pass of radio frequency interference (RFI) excision from the calibrator and target scans is performed using a sliding time median filter. Following these steps, the scripts perform delay and bandpass calibration using the primary calibrators. Time dependent antenna-based complex gain corrections are then derived using the secondary calibrator and interpolated for application to the target scans. A gain correction is derived independently for each spectral window (SPW). The data were processed at the native time and frequency resolution to minimize the effects of smearing away from the phase center. Hanning smoothing was disabled in the VLA pipeline in order to minimise the effects of bandwidth smearing.

Following the execution of the pipeline, SPWs with anomalously high amplitudes were identified and discarded outright. The target field from each pointing was split into a single Measurement Set. The {\textsc CASA} {\tt mstransform} task was then used to add a {\tt WEIGHT\_SPECTRUM} column to the visibilities, and the {\tt statwt} task was used to adjust values in the {\tt WEIGHT\_SPECTRUM} based on the statistical properties of the visibilities for each baseline. Self-calibration did not significantly improve the noise floor in the images. The brighter sources exhibiting the worst artefacts are off-axis, and likely limited by direction-dependent effects, which we did not attempt to correct for.

The target fields were imaged using the {\tt wsclean} software \citep{offringa14}, with the image properties summarised in Table \ref{tab:images}. Images were produced for each band and each cluster by jointly gridding and deconvolving all of the corresponding Measurement Sets. Large image sizes (16,384 $\times$ 16,384 pixels, with pixel scales listed in Table \ref{tab:images}) were used in order to deconvolve sources in the sidelobes of the antenna primary beam to prevent the sidelobes of the synthesised beam associated with them from affecting the target area. A \citet{briggs95} robust parameter of 0.3 was used for all imaging, with additional Gaussian tapers applied to the visibilities as needed.

Spectral behaviour of the sources (both intrinsic towards the beam center, and instrumentally-perturbed off-axis) was captured during deconvolution by imaging the data in four equal spectral chunks across the band. The approach used by {\tt wsclean} during deconvolution is to find peaks in the full band image and then deconvolve these in each sub-band independently. For major-cycle purposes clean components were fitted by a 2nd order polynomial when predicting the visibility model. Cleaning was terminated after 100,000 iterations, which from examination of the model and residual images was deemed to be sufficient without overcleaning. Imaging concludes with the model being restored into the full-band residual map, using a 2D Gaussian as fitted to the main lobe of the point spread function as the restoring beam. Corrections for the conversion of brightness and flux density measurements from apparent to intrinsic values was done via the models derived from holographic measurements of the VLA primary beam by \citet{perley16}.

\section{Data products}
\label{sec:results}

Here we describe the radio images produced for each cluster, as well as the construction of the source catalogs and derivative images, and the verification of these products.

\subsection{Radio images}
\label{sec:images}

\begin{figure*}[ht!]
\centering
\includegraphics[width=0.75 \textwidth]{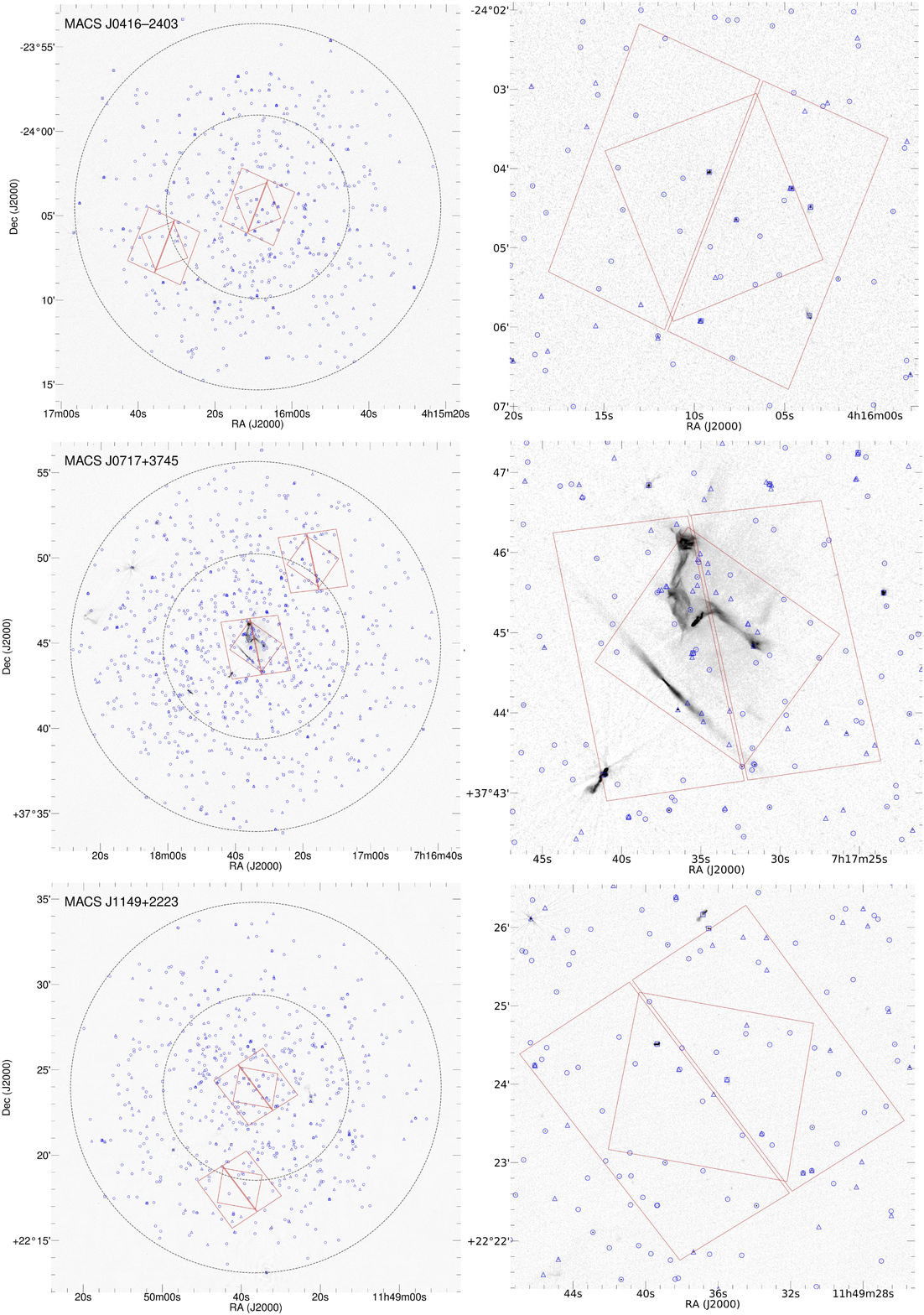}
\caption{The S-HIGH images for the three clusters. The left-hand panels show the full field of view of the S-band observations, with the outer and inner circles showing the nominal cut off points (at the 30\% primary beam level) for the S- and C-band images respectively. The greyscale runs from 0.0 (white) to 40 $\mu$Jy beam$^{-1}$ (black) with a square-root transfer function. The inner red polygon shows the \emph{HST} coverage of the cluster, and the corresponding outer marker shows the \emph{HST} flanking field. The right-hand panels in the figure zoom in on the \emph{HST} cluster area. The blue markers on the figure show cataloged radio sources. Circular markers represent compact radio sources that have a host galaxy identified in either the \emph{HST} or Subaru data. Triangular markers represent compact radio sources for which a cataloged optical / near-infrared host has not been identified. Square markers represent extended morphology radio sources, plotted at the location of the cataloged counterpart from the \emph{HST} or Subaru observations. Please refer to Section \ref{sec:optical} for details of the cross-matching process. \label{fig:maps}}
\end{figure*}

\begin{figure*}[ht!]
\centering
\includegraphics[width=\textwidth]{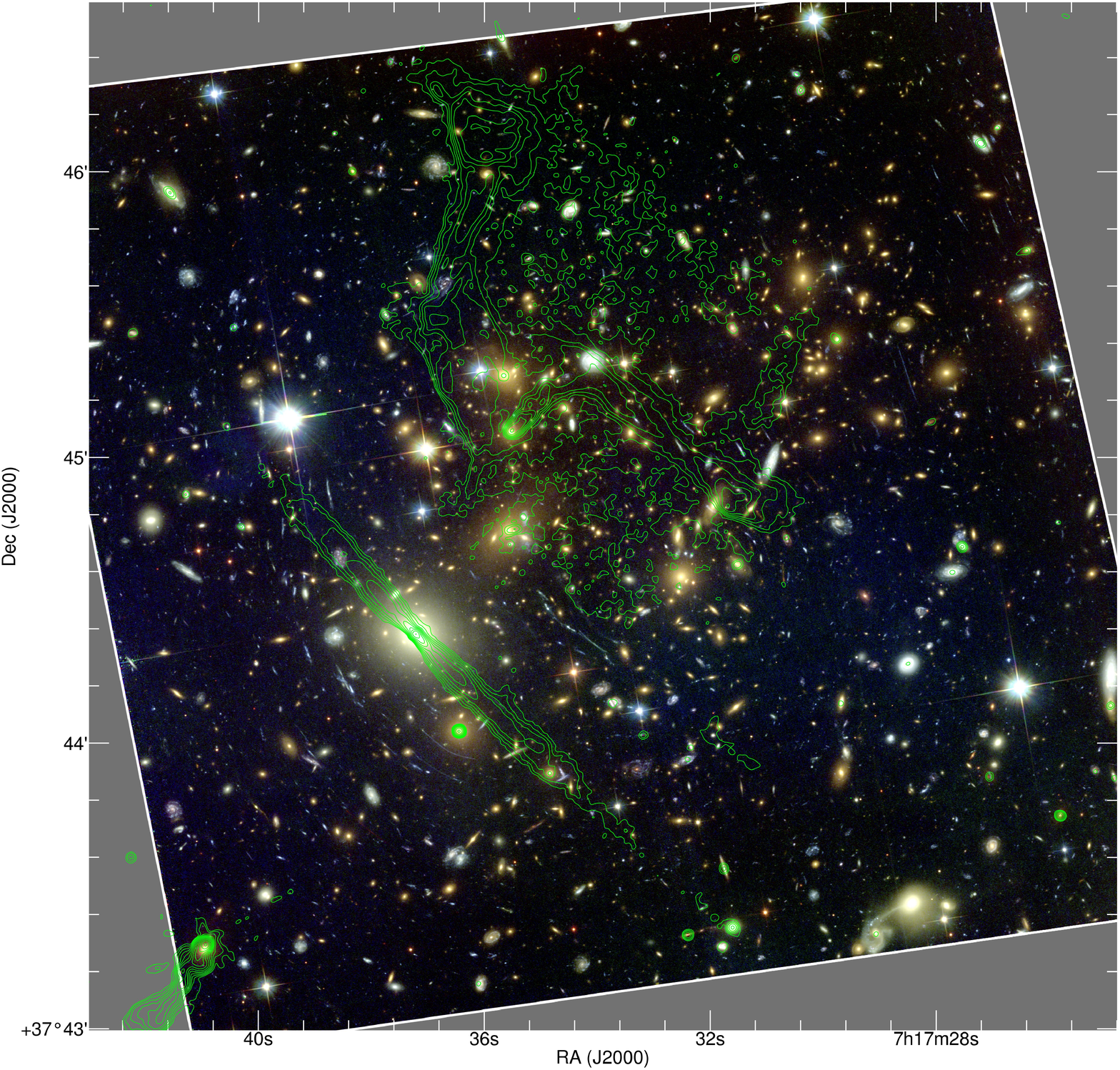}
\caption{The S-HIGH image for MACSJ0717, shown as a contour plot overlaid on a three-colour image formed from the F814W, F606W, and F435W filters from the \emph{HST} observations. The optical counterparts to numerous compact and extended radio sources are visible. The extended structures associated with the bright radio relic in this cluster dominate the radio emission. The base contour level is 1.6 $\mu$Jy~beam$^{-1}$, and from there the contour increments are 1.2~$\times$~$\sqrt{3}^{n}$~$\mu$Jy~beam$^{1}$ where ($n$~=~0,~1,~2,~3$\ldots$).
\label{fig:vlahst}}
\end{figure*}

Four images were produced for each cluster, twelve in total. These are a high-resolution S-band image (S-HIGH), a second S-band image tapered to provide an approximately 3$''$ synthesized beam (S-LOW), a full-resolution C-band image (C-HIGH), and finally, a lower resolution C-band image (C-LOW) designed to match the angular resolution of the S-HIGH image. This is mainly for the determination of matched-resolution component flux densities and brightnesses, and subsequently the determination of component spectral indices. The main properties of each image are summarized in Table \ref{tab:images}. The S-HIGH images are shown in Figure \ref{fig:maps}, and Figure \ref{fig:vlahst} shows contours of the S-HIGH image overlaid on a three-colour \emph{HST} image for MACSJ0717. Further details about the optical catalogs are provided in Section \ref{sec:optical}.

\begin{table*}
\begin{minipage}{175mm}
\centering
\caption{Summary of the image products produced for each field. The angular resolution is the major axis, minor axis, and position angle (east of north) of the two-dimensional Gaussian restoring beam used following deconvolution, in units of arcsec, arcsec, and degrees respectively. The rms noise values are entirely consistent with the expected thermal noise values for these observations.}
\begin{tabular}{llcccc} \hline \hline
Cluster   & Image     & Angular Resolution & Pixel Scale  & rms noise \\ 
          &           & [$''$, $''$, $^{\circ}$]  & [$''$]       & [$\mu$Jy beam$^{-1}$]    \\ \hline
MACSJ0416 & S-HIGH    & 0.94, 0.51, 1.9   & 0.16         & 1.0            \\
          & S-LOW     & 4.39, 3.12, 43.2  & 0.5          & 1.8            \\	
          & C-HIGH    & 0.53, 0.3, 22.7   & 0.05         & 0.9            \\
          & C-LOW     & 0.94, 0.68, 27.7  & 0.12         & 1.0            \\
MACSJ0717 & S-HIGH    & 0.73, 0.61, 93.5  & 0.16         & 0.7            \\
          & S-LOW     & 3.57, 3.42, 79.0  & 0.5          & 1.2            \\
          & C-HIGH    & 0.33, 0.27, 112.7 & 0.05         & 1.0            \\
          & C-LOW     & 0.74, 0.69, 70.8  & 0.12         & 1.1            \\
MACSJ1149 & S-HIGH    & 0.51, 0.48, 35.9  & 0.16         & 0.9            \\
          & S-LOW     & 3.16, 3.05, 101.8 & 0.5          & 1.6           \\
          & C-HIGH    & 0.28, 0.27, 70.9  & 0.05         & 0.9            \\
          & C-LOW     & 0.7, 0.65, 100.5  & 0.12         & 1.0            \\ \hline
\end{tabular}
\label{tab:images}
\end{minipage}
\end{table*}

\subsection{Generation of the compact radio component catalogs}
\label{sec:compactcat}

The `master' map for component extraction is taken to be the S-HIGH image for each cluster, as this provides both a higher surface density of detections and a larger area compared to the C-band images. The first step involved running the PyBDSF \citep{mohan15} source finder on this image using a peak threshold of 5$\sigma$ and an island threshold of 3$\sigma$, where $\sigma$ is taken to be the source-finder's own estimate of the background noise level. This is determined by stepping a box across the image and computing the standard deviation of the pixels within that region, and the resulting position-dependent values are recorded and interpolated. This produces a background noise image that is used for local estimates of $\sigma$. Peaks above 5 times this value are identified, and then a flood-fill method is used to delineate regions of contiguous emission down to the secondary 3$\sigma$ island threshold. Regions of significant emission are then decomposed into a series of points and Gaussian components. Default box size settings were used in each case. This procedure was performed on each of the three S-HIGH images.	

Following this, the resulting catalogs were manually examined to remove spurious components from image artefacts. These were primarily associated with brighter sources away from the pointing center. Components belonging to sources with extended morphologies were also visually identified and placed into a separate catalog (see Section \ref{sec:extended} and Appendix \ref{sec:extendedappendix}). Following these steps, we obtain a catalog containing only the compact S-band radio sources for each cluster.

\subsection{Identifying optical / near-infrared hosts}
\label{sec:optical}

Determining optical identifications for the radio sources relies on three existing resources for each cluster. The first two of these is a set of \emph{HST} photometric catalogs created by the ASTRODEEP\footnote{{\tt http://www.astrodeep.eu}} project, by \citet{castellano16} for MACSJ0416, and by \citet{dicriscienzo17} for MACSJ0717 and MACSJ1149, as well as the more recent catalogs from the DeepSpace project \citep{shipley18}. These catalogs also include spectroscopic redshifts where available, from the observing campaign by \citet{ebeling14}. The third resource consists of the source catalogs derived from wider-area imaging with the Subaru telescope by the CLASH team, complete with photometric redshifts as presented by \citet{medezinski13} for MACSJ0717 and \citet{umetsu14} for MACSJ0416 and MACSJ1149.

A cutout image around each compact component was created. The radio contours were overlaid on either the \emph{HST} F140W image or the Subaru $z$-band image if the region was outside the \emph{HST} footprint. Entries from either the ASTRODEEP \emph{HST} or Subaru catalogs were overlaid on the cutout image, and visual inspection determined whether a cataloged optical / near-infrared association for the radio component was present or not, based on positional and morphological considerations. Note that we validate the astrometric alignment between radio and optical in Appendix \ref{sec:astrometry}, and the cross-matching process is unaffected by this. In addition to verifying potential hosts for the radio components, manually inspecting each of the compact components also allowed flags to be applied to sources that: (i) could be well-represented by single Gaussian components but had been decomposed into several components by {\tt PyBDSF}; and (ii) sources that had significantly extended structures associated with them that had been missed in the first pass. Components belonging to the first category were force-fit with a single Gaussian using the {\tt CASA imfit} task, and the multiple entries belonging to this source in the component catalog were replaced by this single component. Radio components belonging to the second category were removed from the compact catalog and placed into the extended catalog. Note that this process only cross-matched radio components with \emph{cataloged} optical sources from the \emph{HST} or Subaru catalogs. 

At this stage the compact S-band catalog for each cluster exists in two subsets: those with optical / near-infrared matches (the `radio-optical' catalog) and those without (the `radio' catalog). Relevant properties derived from the \emph{HST} observations (e.g. optical IDs, positions, stellar mass and star formation rate estimates, spectroscopic and photometric redshifts) are merged with the radio-optical catalog, using the more recent catalogs from the DeepSpace project \citep{shipley18}, using positional coincidence matches between these catalogs and the ASTRODEEP catalogs. Additional entries present in the DeepSpace catalogs that were not visually matched as part of the ASTRODEEP process were assigned to radio components by conducting nearest-neighbour crossmatching for the additional sources that the DeepSpace catalogs contain. Following e.g.~\citet{ivison07}, the match radius was set according to
\begin{equation}
\label{eq:pos_scatter}
\Delta_{\mathrm{RA}}~=~\Delta_{\mathrm{Dec}}~=~0.66 \frac{\theta}{\mathrm{SNR}}
\end{equation}
which predicts how $\Delta_{\mathrm{RA}}$ and $\Delta_{\mathrm{Dec}}$ (the scatter in right ascension and declination of the radio component) scale with the signal-to-noise ratio (SNR) of the detection for a synthesised beam with a FWHM of $\theta$. Evaluating Equation \ref{eq:pos_scatter} for SNR~=~5 for the S-HIGH images (according to the noise values reported in Table \ref{tab:images}), results in match radii of 0.22$''$, 0.16$''$, and 0.12$''$ for MACSJ0416, MACSJ0717 and MACSJ1149 respectively. Note that prior to cross-matching the offsets described in Appendix \ref{sec:astrometry} were removed to further align the cataloged positions. A complete description of the columns for both the radio-optical and the radio catalog, including the properties derived in the sections that follow, is provided in Appendix \ref{sec:radoptstructure}. Stellar mass, star-formation rates, and specific star-formation rates from the DeepSpace are included in the radio-optical catalog where appropriate. For a detailed analysis of radio vs optical star-formation rates we refer the reader to the companion paper to this work by Jim\'{e}nez-Andrade et al. (\emph{submitted}).

\begin{figure}[h]
\centering
\includegraphics[width=\columnwidth]{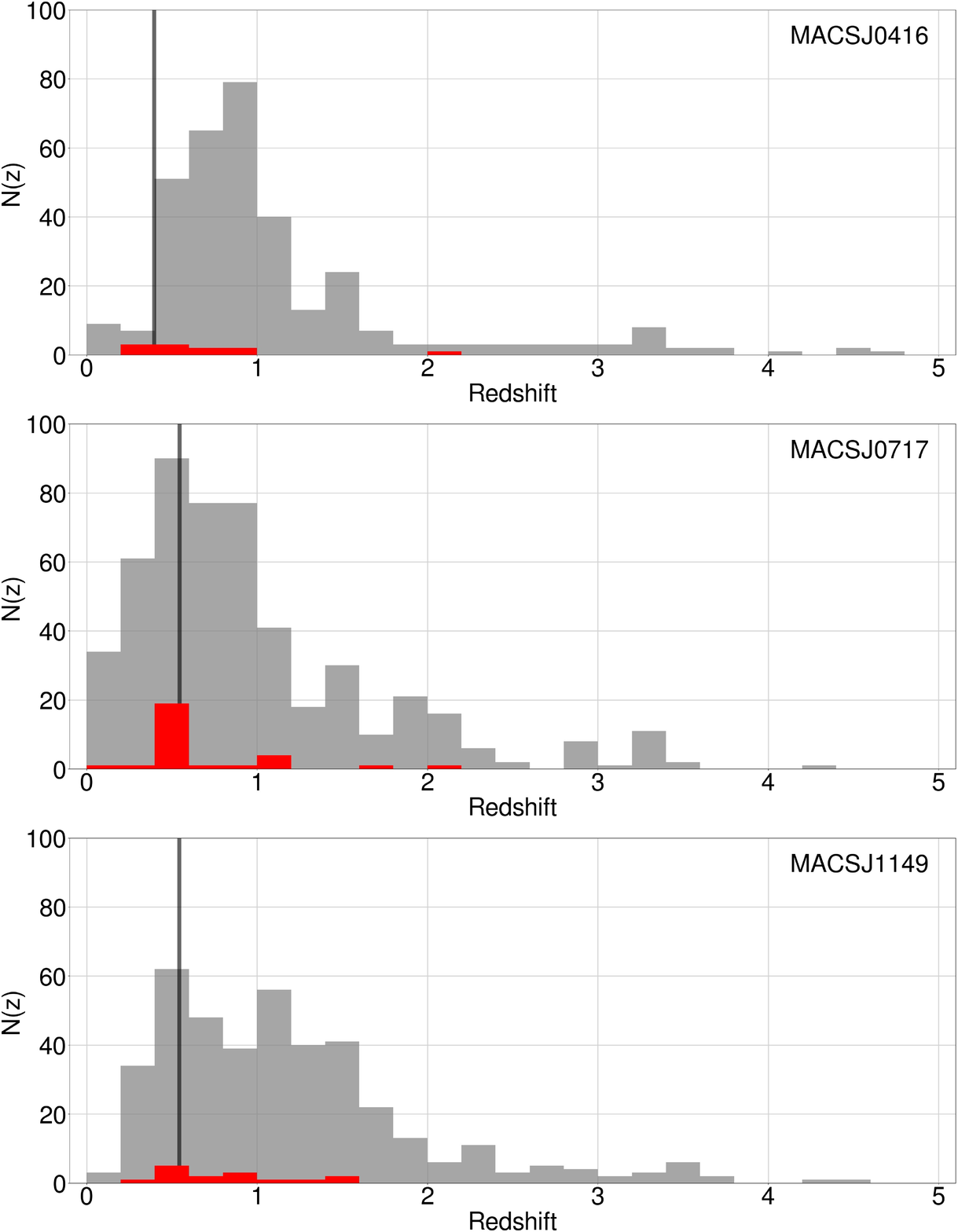}
\caption{Distribution of the spectroscopic (red) and photometric (grey) redshifts as measured from the optical IDs of the radio components detected in this survey. This includes the host galaxies of extended radio structures. The median redshift measured across all three clusters is 0.88 The vertical lines on each plot mark the redshifts of the clusters.}
\label{fig:z_dist}
\end{figure}

\subsection{Resolved sources, uncertainties in the deconvolved sizes, and the ``best'' flux density estimates}
\label{sec:s-star}

Although {\tt PyBDSF} provides the deconvolved source sizes, the associated uncertainties in the major and minor axis, and position angle, that are returned are the same as the apparent sizes in the map plane (at the time of writing). We thus estimate intrinsic source (major axis) sizes and the associated uncertainties according to the method presented in Appendix \ref{sec:sizes}.

In addition to the component peak brightnesses and integrated flux densities measured by the source finder, we also provide a `best' flux density estimate ($S_{*}$) based on an assessment of how reliably resolved the Gaussian components are. This largely follows the method adopted by \citet{murphy17}, modified for the considerations above that cater for the fact that the restoring beam in our maps was not forced to be circular. If the criterion
\begin{equation}
\phi^{'} - \theta^{'}_{\mathrm{beam}}~\geq~2\sigma^{'}_{\phi}
\label{eq:resolved}
\end{equation}
is satisfied then we deem a source to be reliably resolved, where $\theta^{'}$ is the FWHM of the elliptical beam as projected along the source major axis, and $\phi^{'}$ and $\sigma^{'}_{\phi}$ are the source major axis and associated uncertainty as measured by {\tt PyBDSF}. In this case, $S_{*}$ is set to the integrated flux density of the component as determined by the source finder. For sources that are marginally resolved (those that do not satisfy the Equation \ref{eq:resolved} criterion) we set $S_{*}$ to be the geometric mean of the integrated flux density and peak brightness. The uncertainties in these two quantities are anti-correlated when fitting Gaussians, and hence the geometric mean provides the best flux density measurement for a Gaussian fit to a faint component \citep{condon97}.

\subsection{Adding 6~GHz information and determining spectral indices}
\label{sec:cband}

The PyBDSF source finder was also run on the C-LOW images, and the results were cross-matched with the `radio-optical' and `radio' catalogs described above. The C-LOW image matches the S-HIGH images in terms of angular resolution, making for more robust flux density comparisons for compact features between the two frequencies, and thus for more reliable spectral index estimates. All components in the S-band catalog that had a C-band match within a radius of 1$''$ were visually examined (via S-band contours over the C-band image) to confirm that the C-band component was not spurious, nor fragmented into multiple components. Genuine C-band matches were then associated with the relevant S-band components. Spectral index ($\alpha$)\footnote{Here we define the flux density $S$ being related to frequency $\nu$) via the spectral index $\alpha$ according to: $S \propto \nu^{\alpha}$.} estimates for compact components were derived from the dual-frequency ``best'' flux density measurements ($S_{*}$), and added to the catalog.

\subsection{Lensing magnifications}
\label{sec:mags}

The Frontier Fields have publicly available lensing models\footnote{https://archive.stsci.edu/prepds/frontier/lensmodels/}, derived independently by five different teams. Further details are given by \citet{lotz17}, with the mapping teams making use of data from \citet{schmidt14}, \citet{vanzella14}, \citet{diego15}, \citet{merlin15}, \citet{castellano16}, \citet{jauzac16}, \citet{kawamata16}, \citet{limousin16}, \citet{merlin16} and \citet{caminha17}.

The models take the form of images that capture the spatial variation in mass surface density ($\kappa$) and lensing shear ($\gamma$) over each cluster. The magnification ($\mu$) for a point-like source at a given redshift ($z_{s}$) behind the lensing cluster (at redshift $z_{l}$) can be modeled using these two parameters via the relationship:
\begin{equation}
1/\mu~=~(1~-~\kappa_{z})^{2}~-~\gamma_{z}^{2},
\label{eq:mu}
\end{equation}
where $\kappa_{z}$ and $\gamma_{z}$ are the model values for mass surface density and shear scaled by the ratio $D_{ls}/D_{s}$, where $D_{ls}$ is the angular diameter distance between the cluster and the source, and $D_{s}$ is the angular diameter distance from redshift zero to the source.

Magnification estimates were determined for radio components that have either spectroscopic or photometric redshifts by evaluating Equation \ref{eq:mu}, using the source redshift estimate. The preferred order of redshifts when calculating lensing is (i) spectroscopic redshifts from the DeepSpace catalogs; (ii) photometric redshifts from the DeepSpace photometry fitting as derived using the EAZY code \citep{brammer08}; (iii) Bayesian photometric redshifts from the CLASH catalogs. This process was repeated for each lensing model, and the median magnification over all available models is evaluated and recorded in the catalog. Median absolute deviations of $\mu$ across all available models are also tabulated in order to provide some measure of the magnification uncertainty. Note that when extracting the values from the maps of $\kappa$ and $\gamma$, the position of the radio component was used, not the position of the optical host, naturally leading to some differences between the magnifications in the radio-optical catalog and those derived from the DeepSpace catalogs. Note that we do not track inverted parity ($\mu$~$<$~0) sources in the catalog as we only present the median absolute value over all available lensing models. 

\subsection{Extended radio sources}
\label{sec:extended}

As mentioned in Section \ref{sec:compactcat}, the visual confirmation of optical hosts in conjunction with the component models derived by PyBDSF allowed a thorough identification of radio sources with extended morphologies. These features were sub-divided into three categories: (i) extended radio sources with cataloged optical hosts; (ii) extended radio sources with no cataloged optical host; (iii) diffuse structures (e.g.~relics and haloes) that are likely arising due to processes in the foreground clusters. Optical IDs were associated with the first category via the same method as for the compact components, and categories (i) and (ii) were distinguished from (iii) using morphological and optical-ID considerations. 

In order to determine the integrated flux density of the extended sources we employ the {\tt ProFound} software package \citep{robotham18}. Designed primarily with large optical surveys in mind, the software has been found to provide superior photometric estimates to other common source finding packages when applied to both simulated and real radio interferometer data \citep{hale19}. {\tt ProFound} does not perform component fits to regions of extended emission, instead adopting an approach whereby thresholded regions are iteratively dilated until the surface brightness measurement converges. 

Cut-out images for each extended region from both the S-LOW and S-HIGH images were processed using {\tt ProFound}, the former also being used to better gauge the total integrated flux density for sources that may be partially resolved-out by the A-configuration observations. The resulting catalogs from ProFound were pruned of any measurements of additional sources in the field, in order to only derive properties from the extended source being considered. 

Optical / near-infrared and radio contour composite images, notes on individual extended sources, and their tabulated radio properties are presented in detail in Appendix \ref{sec:extendedappendix}. Note that the well-studied foreground FR-I source in the MACSJ0717 field, and the narrow-angle tail galaxy embedded in the radio relic/halo are not included.

\section{Results}
\label{sec:discussion}

\subsection{Redshift distributions of radio sources}
\label{sec:redshifts}

There are 1,262 / 55 radio sources with a photometric / spectroscopic redshift across the three fields, including the host galaxies of radio sources with complex morphologies. The distribution of these redshifts is shown in Figure \ref{fig:z_dist}, with photometric measurements shown in grey and spectroscopic measurements shown in red. The median redshift measured from the catalogd sources from all three clusters is 0.88. This is lower than generally reported from other radio-selected deep-field studies \citep[e.g.][]{smolcic17}, however the over-densities of galaxies in the clusters themselves will pull this median value down. On a per-cluster basis, the median redshifts are 0.9, 0.75, and 1.03 for MACSJ0416, MACSJ0717, and MACSJ1149 respectively. If we exclude sources within $\pm$0.05 of the cluster redshift, the median redshifts of the radio-selected objects become 0.92, 0.85, and 1.13, having excluded 10, 58, and 38 objects from MACSJ0416, MACSJ0717, and MACSJ1149 respectively. There appears to be significantly more sources out to $z$~$\sim$~1.5 in MACSJ1149, and the detections in this field appear to be less clustered along the line of sight (see also Figure \ref{fig:radioluminosities}). Despite the limiting magnitudes for the Subaru observations being the same, the surface density of strongly-lensed sources varies by a factor $>$2 across the three fields \citep{umetsu14}. This suggests that sample variance is the likely explanation for the boxy (or possibly bimodal) redshift distribution of the radio sources in MACSJ1149.

\subsection{Spectral indices of compact radio sources}
\label{sec:alphas}

Accurate radio spectral index ($\alpha$) measurements are essential for deriving rest-frame quantities such as spectral luminosity, and by extension parameters such as star-formation rates (e.g.~Jim\'{e}nez-Andrade et al., \emph{submitted}). The use of measured values of $\alpha$ as opposed to adopting charactersitic or canonical values (typically assumed to be $-$0.7 to $-$0.8 for synchrotron emission) can significantly reduce the scatter and biases in studies involving the radio / far-infrared correlation, as demonstrated by \citet{gim19}, who also demonstrate the importance of deriving values of $\alpha$ from dual-frequency images that are resolution-matched.

\begin{figure}[h]
\centering
\includegraphics[width=\columnwidth]{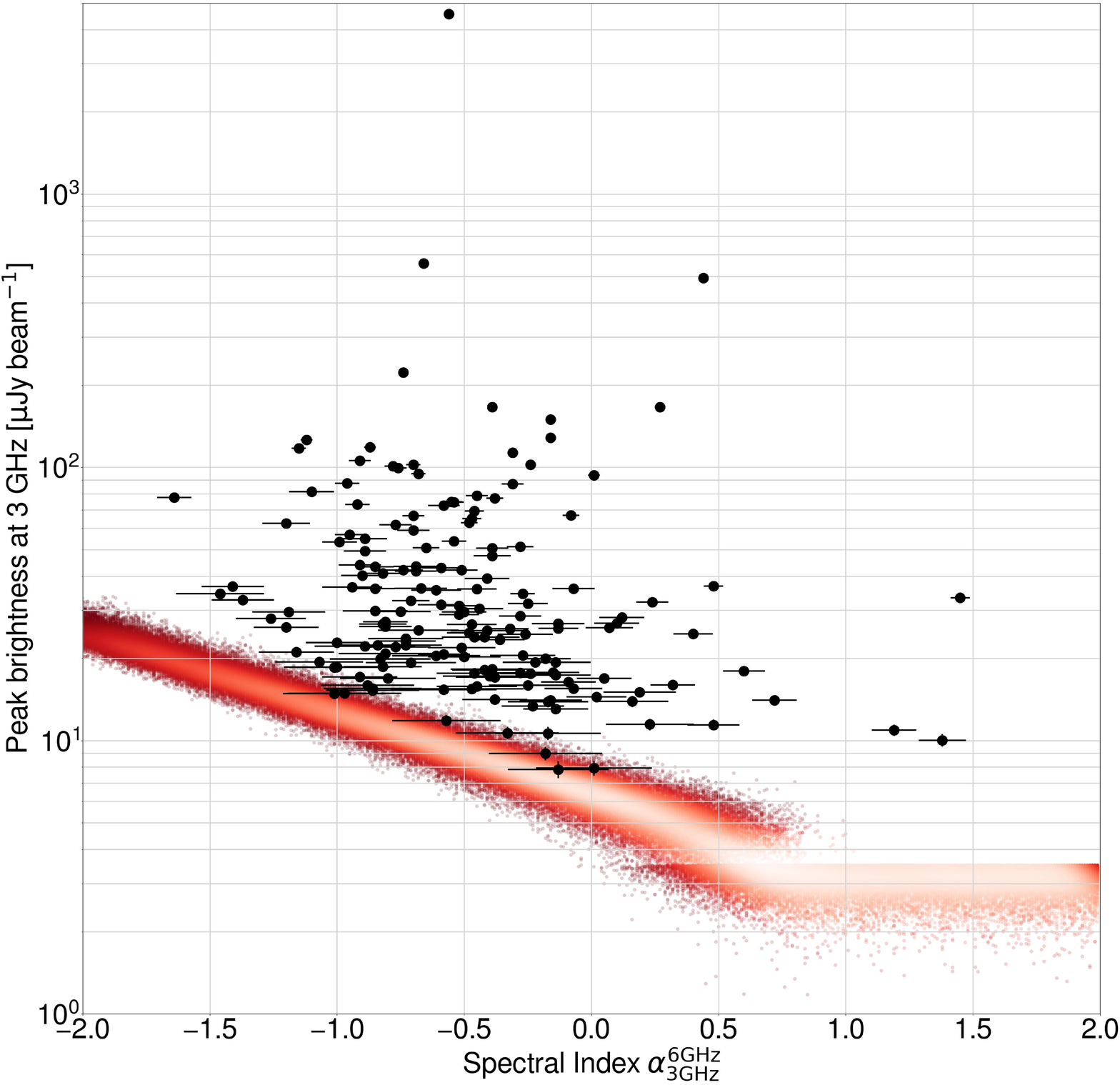}
\caption{Integrated flux densities of the S-band radio components against their 3-6 GHz spectral index value. The cloud of points shows the results of a Monte Carlo simulation used to gauge the selection function in this plane. Note that this is a population of simulated sources that are \emph{non-}detections. Sources below this distribution should be undetectable, and our catalog is only complete for -2~$<$~$\alpha$~$<$~2 for peak 3~GHz brightnesses above 30 $\mu$Jy~beam$^{-1}$. Please see Section \ref{sec:alphas} for details.}
\label{fig:alphas}
\end{figure}

Figure \ref{fig:alphas} shows the 3~GHz integrated flux densities of the compact radio sources against their 3--6 GHz spectral index. Note that the plot only shows sources that have a C-band detection within the 50\% gain region of the VLA C-band primary beam at 6~GHz (for a total of 169 objects). Within this region both the C-band and especially S-band primary beam responses do not impart significant attenuation, and thus the selection biases introduced by the frequency-dependent antenna primary beam pattern over the broad bandwidths of both observing bands are reduced. Selection biases persist however, introduced by the unknown distribution of source spectral indices and the slightly different depths of the S-HIGH and C-LOW images, which are used to measure $\alpha$ for the cataloged sources.

We investigate the selection function imposed on the S-$\alpha$ plot using a Monte Carlo simulation. The spectral index range -2 to +2 is partitioned into 400 bins (width 0.01). For each of these bins a source with the corresponding spectral index is created and assigned a peak brightness at 3~GHz. The corresponding peak brightness at 6~GHz is calculated according to the assigned value of $\alpha$. Both the 3 and 6 GHz values are then perturbed with noise drawn from a Gaussian distribution with a mean of zero and a standard deviation corresponding to the lowest S-HIGH (0.7 $\mu$Jy beam$^{-1}$) and C-LOW (1.0 $\mu$Jy beam$^{-1}$) rms noise values across the three clusters, as listed in Table \ref{tab:images}. If the noisy simulated brightness measurements at 3 and 6 GHz exceed five times the rms noise values in both bands, then the simulated source is considered to be detected, and in the real-world case would have a legitimate $\alpha$ measurement in the catalog. Since we are interested in the sources we may be missing from Figure \ref{fig:alphas}, the above procedure is repeated with a steadily declining 3 GHz brightness. When the simulated source does not meet the peak brightness detection criteria, i.e. either the 3 or 6 GHz brightness drops below the threshold, the source is considered undetected and the peak brightness and $\alpha$ values are noted. This process is repeated 400 times for all 400 $\alpha$ bins and the resulting distribution of simulated sources that have dropped below the detection threshold is plotted as the red cloud of points on Figure \ref{fig:alphas}, where the density of sources increases from dark red to white.

The fact that there is a detection criterion at both 3 and 6 GHz, but the plot shows only the 3 GHz measurements, imparts two distinct regions to this distribution. Sources to the left of the distribution have steep spectra that render them too faint for detection at 6 GHz, but the corresponding full range of noisy 3 GHz brightness measurements is present. Sources to the right of the distribution have inverted spectra that cause them to drop below the 3 GHz detection threshold, delineated by the hard upper limit on the right of the distribution. Essentially we are seeing the manifestation of Eddington bias as it applies to a combination of two peak brightness-limited samples. Our spectral index selection function is thus only complete for -2~$<$~$\alpha$~$<$~2 above a peak 3~GHz brightness of 30 $\mu$Jy~beam$^{-1}$.

The median spectral index of the 169 objects plotted in Figure \ref{fig:alphas} is -0.5, however for the 74 sources with peak brightnesses above 30 $\mu$Jy~beam$^{-1}$ at 3~GHz the median value of $\alpha$ is -0.63. For the canonical synchrotron radiation spectral index of -0.7, our peak S-band completeness limit is equal to 50 $\mu$Jy~beam$^{-1}$ at 1.4~GHz. Our median value is consistent with previous dual-band studies at comparable depths \citep[e.g.][]{huynh15, gim19, huynh20}, as well as in-band measurements made using the 1--2 GHz L-band receivers of the VLA \citep{heywood16, heywood20}. We note also that the dual-band studies involving measurements at $\sim$5~GHz and above also revealed a significant fraction of flat and inverted spectrum sources in addition to those with typical synchrotron spectra, something that is also evident in Figure \ref{fig:alphas}.

\subsection{Lensed compact radio sources}
\label{sec:lensedsources}

\begin{figure}[h]
\centering
\includegraphics[width=\columnwidth]{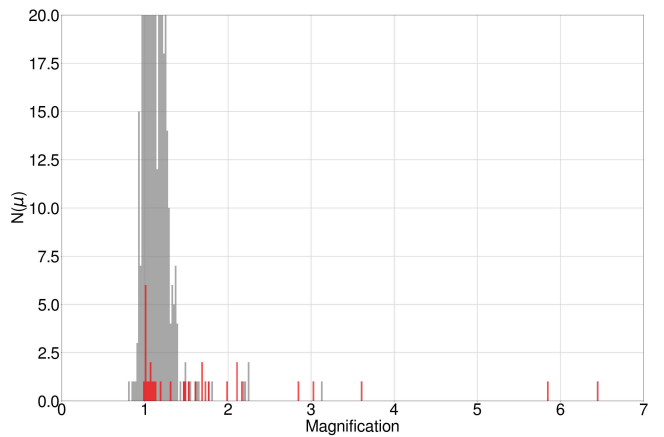}
\caption{Distribution of the lensing magnifications as derived in Section \ref{sec:mags} for sources with spectroscopic redshifts (red) and photometric redshifts (grey). The bin widths are 0.2. The bins in the 1.0 - 1.2 range run over the figure limit, dominated by those for which a unity magnification is entered into the catalog as they are beyond the regions covered by the lensing models. Several sources are demagnified (0$<$$\mu$~$<$~1). Note that we do not track inverted parity ($\mu$~$<$~0) sources.}
\label{fig:mags}
\end{figure}

\begin{figure*}[ht!]
\centering
\includegraphics[width=\textwidth]{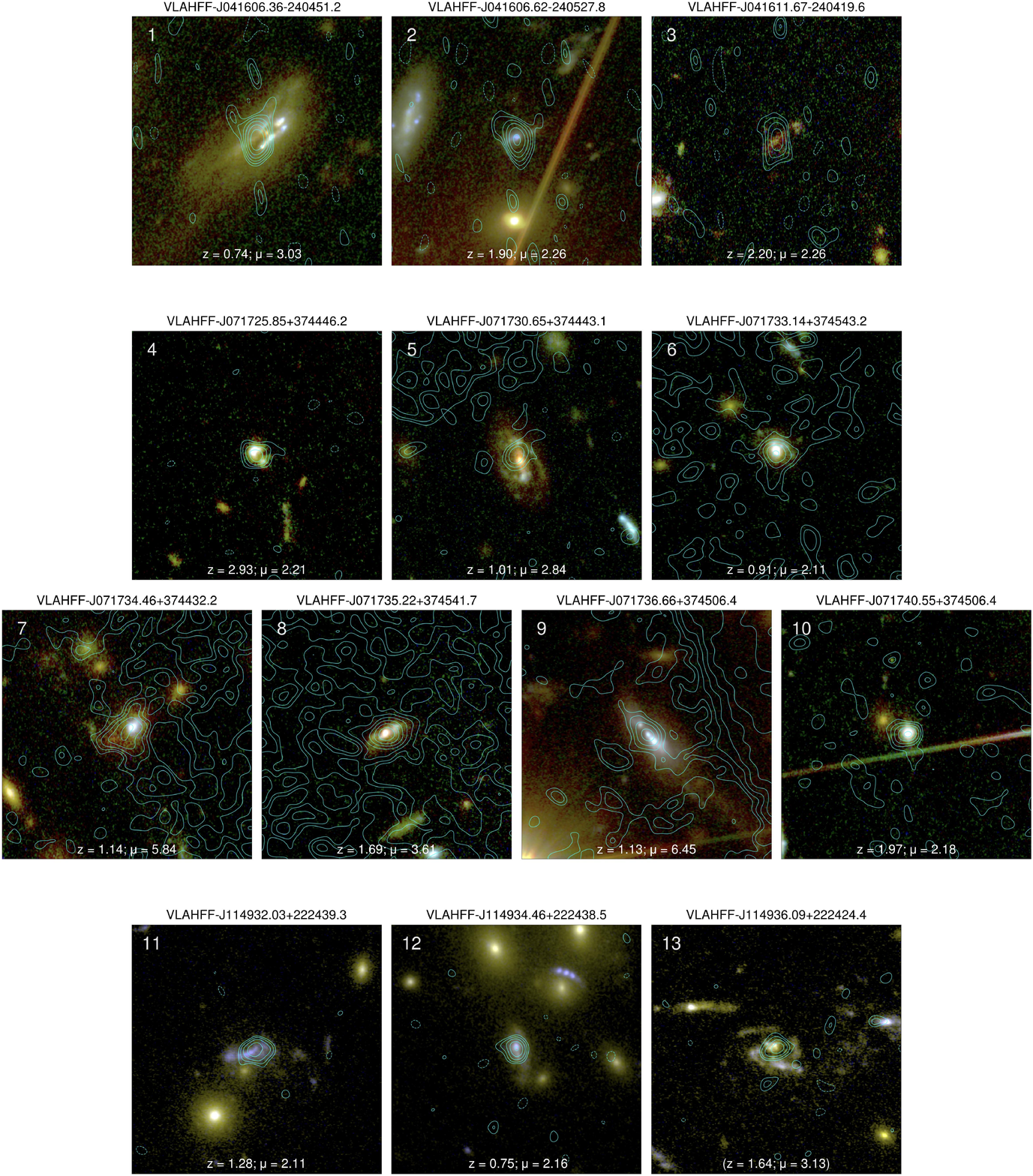}
\caption{Radio contours over \emph{HST} images showing the compact radio sources with lensing magnifications greater than 2. Panels 1--3 show sources behind MACSJ0416, 4--10 show sources behind MACSJ0717, and 11--13 show sources behind MACS1149. The \emph{HST} RGB images are derived from the F814W, F606W and F435W filters respectively. Each panel spans 8$''$.64. The contours trace the S-HIGH image, with levels at 1.7 $\times$ (1, $\sqrt{2}$, 2, 2$\sqrt{2}$, 4, 4$\sqrt{2}$, 8, 8$\sqrt{2}$, \ldots) $\mu$Jy beam$^{-1}$, with a single (dashed line) negative contour at $-$1.7 $\mu$Jy beam$^{-1}$. Note that the crowded contours in some of the MACSJ0717 panels are due to the cluster relic, and not elevated noise levels. Source redshifts and median magnification values are indicated on each panel. Demagnified integrated flux densities, peak brightnesses, and effective noise levels for each source are listed in Table \ref{tab:demagnified}.\\ \label{fig:lensed_source_cutouts}}
\end{figure*}

Figure \ref{fig:mags} shows a histogram of the lensing magnification estimates, the derivation of which is given in Section \ref{sec:mags}. As with Figure \ref{fig:z_dist}, sources with spectroscopic redshifts are marked in red. We detect 13 objects with magnification factors greater than 2, seven of which have spectroscopic redshifts. Figure \ref{fig:lensed_source_cutouts} shows cutouts for each of these sources. The ID, redshift, and median lensing magnification is given above and on the numbered panel for each source. The background of each image is a three-colour rendering of the F814W, F606W, and F435W \emph{HST} filters overlaid with the S-HIGH contours, with contour levels provided in the caption.

The majority of these sources are blue star-forming galaxies with spiral or disturbed morphologies and prominent dust lanes. The mean redshift is 1.73 (standard deviation 0.81). For each of the radio components shown in Figure \ref{fig:lensed_source_cutouts} we provide demagnified, intrinsic values for the integrated flux density, peak brightness, and effective noise levels in Table \ref{tab:demagnified}. The source with the highest magnification factor ($\mu_{\mathrm{median}}$~=~6.45) is VLAHFF-J071736.66+374506.4, which has an intrinsic peak brightness of 0.9 
$\mu$Jy beam$^{-1}$ at an effective noise level of 140 nJy beam$^{-1}$. This makes it a candidate for the faintest radio source detected to date \citep[see also][]{jackson11}. The total star formation rate for this source is modest, at 10 M$_{\odot}$ year$^{-1}$ (for more details see Jim\'{e}nez-Andrade et al., \emph{submitted}).

\begin{table*}
\begin{minipage}{175mm}
\centering
\caption{Magnifications, and demagnified integrated flux densities, peak brightnesses, and effective noise levels  for the 13 lensed sources presented in Figure \ref{fig:lensed_source_cutouts}. The first column refers to the panel number for each source in that figure.}
\begin{tabular}{llcccc} \hline \hline
Panel & ID & $\mu$ & $S^{\mathrm{demagnified}}_{\mathrm{int}}$ & $S^{\mathrm{demagnified}}_{\mathrm{peak}}$ & $\sigma^{\mathrm{demagnified}}$ \\
 & & & [$\mu$Jy] & [$\mu$Jy beam$^{-1}$] & [$\mu$Jy beam$^{-1}$] \\ \hline
1  & VLAHFF-J041606.36$-$240451.2 & 3.03 & 6.71 & 4.76 & 0.33 \\
2  & VLAHFF-J041606.62$-$240527.8 & 2.26 & 7.75 & 5.06 & 0.44 \\
3  & VLAHFF-J041611.67$-$240419.6 & 2.26 & 5.75 & 2.91 & 0.44 \\
4  & VLAHFF-J071725.85+374446.2 & 2.21 & 4.04 & 3.03 & 0.41 \\
5  & VLAHFF-J071730.65+374443.1 & 2.84 & 2.44 & 1.85 & 0.32 \\
6  & VLAHFF-J071733.14+374543.2 & 2.11 & 8.43 & 3.1  & 0.43 \\
7  & VLAHFF-J071734.46+374432.2 & 5.84 & 7.57 & 1.14 & 0.15 \\
8  & VLAHFF-J071735.22+374541.7 & 3.61 & 4.61 & 4.56 & 0.25 \\
9  & VLAHFF-J071736.66+374506.4 & 6.45 & 1.96 & 0.9  & 0.14 \\
10 & VLAHFF-J071740.55+374506.4 & 2.18 & 4.17 & 3.78 & 0.41 \\
11 & VLAHFF-J114932.03+222439.3 & 2.11 & 9.99 & 3.04 & 0.43 \\
12 & VLAHFF-J114934.46+222438.5 & 2.16 & 6.05 & 3.61 & 0.42 \\
13 & VLAHFF-J114936.09+222424.4 & 3.13 & 4.26 & 2.11 & 0.29 \\ \hline
\end{tabular}
\label{tab:demagnified}
\end{minipage}
\end{table*}

We note some differences between our lensed source catalog and that of \citet{vanweeren16}. The principal reason for this is our use of median rather than mean magnifications. The former is far more robust to strong outliers that may be present in the range of lensing models used. For example, when using mean magnifications for our catalogs one source had an average magnification factor of 150. Examination of the nine individual magnification values from each of the models revealed one model predicting a magnification factor of 1305.8, whereas the median of the eight other values was 6.22. The use of median values also results in fewer sources with $\mu$ $>$ 2.	

\subsection{Demagnified 3~GHz radio luminosities, and the discovery of a powerful radio galaxy at $z~>~4$}

\begin{figure}[h]
\centering
\includegraphics[width=\columnwidth]{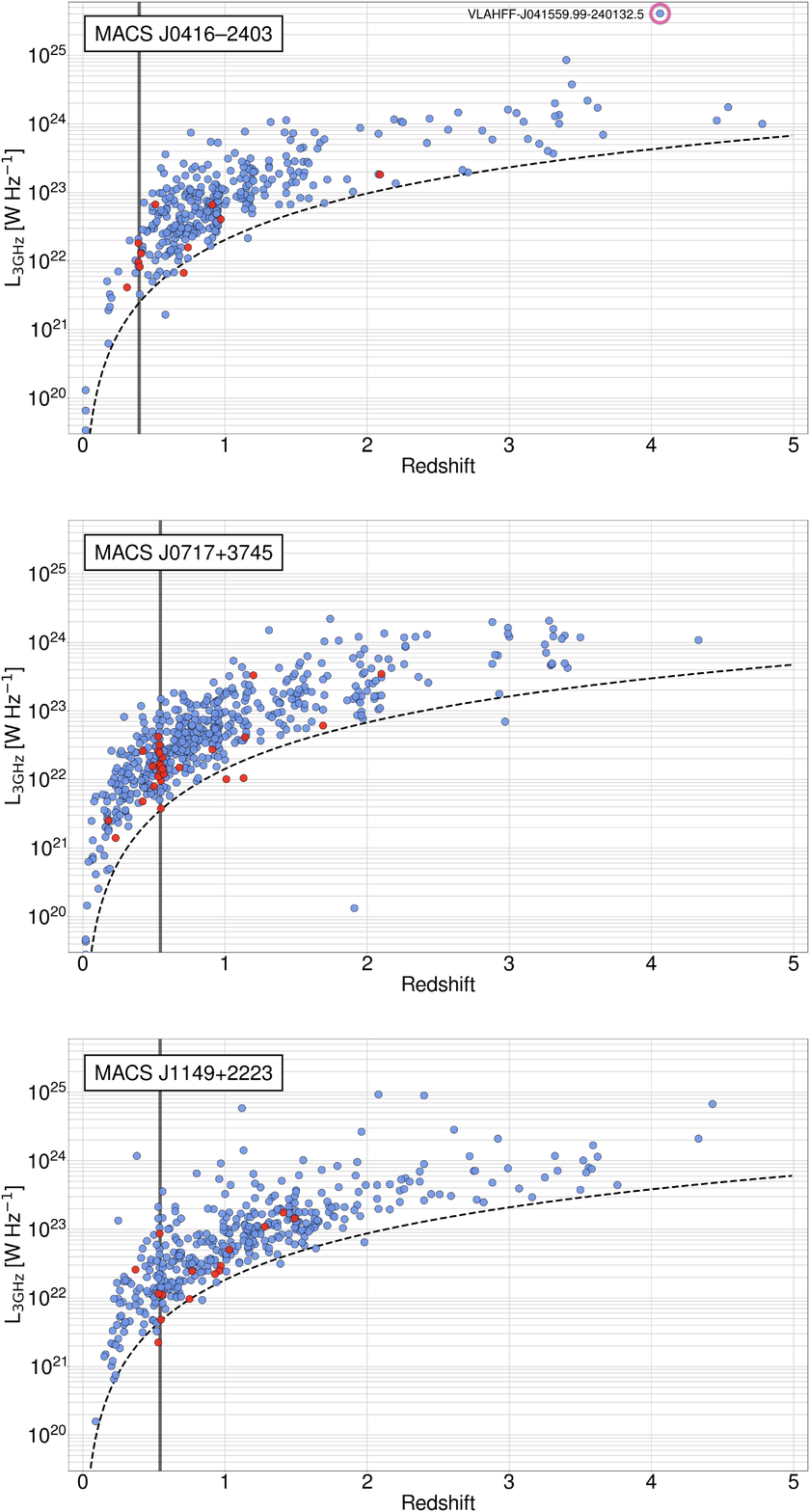}
\caption{The 3~GHz radio luminosities of the cataloged radio sources plotted against redshift for each of the three clusters. These have been demagnified according to the lensing magnification estimates in order to provide intrinsic luminosities. Sources with spectroscopic redshifts are marked in red. The vertical lines show the cluster redshifts. The dashed line shows the 5$\sigma$ detection limit based on the per-cluster S-HIGH noise measurements as provided in Table \ref{tab:images}. Note that the extreme outlier in the MACSJ0717 field is likely the result of an improper photometric redshift fit. The highlighed source is likely a newly-discovered powerful radio galaxy at z~=~4.06.}
\label{fig:radioluminosities}
\end{figure}

Here we bring together the results of Sections \ref{sec:redshifts}, \ref{sec:alphas}, and \ref{sec:lensedsources} to compute the demagnified 3~GHz radio luminosities for the compact radio components. These are plotted against redshift for each cluster in Figure \ref{fig:radioluminosities}. The vertical lines show the redshifts of the foreground clusters on each panel, and the dashed line shows the 5$\sigma$ detection limit based on the noise measurements of the S-HIGH images as listed in Table \ref{tab:images}. Measured spectral index values are used where available, and our median value of $-$0.63 is used otherwise. The preferential order for redshifts is as before, namely (i) DeepSpace spectroscopic redshifts; (ii) DeepSpace photometric redshifts; (iii) CLASH  photometric redshifts. As elsewhere, the latter dominate the counts. The magnification corrections demonstrate that we are detecting high redshift galaxies below the formal detection threshold, although the extreme outlier visible in the MACSJ0717 field is likely the result of an improper photometric redshift fit. Figure \ref{fig:radioluminosities} also shows evidence for more distant clustering in redshift along the line of sight, particularly in the MACSJ0416 field.

The MACSJ0416 luminosity plot reveals a source at $z$~=~4.06 (VLAHFF-J041559.99-240132.5), with a 3~GHz  rest-frame luminosity of 4.1~$\times$~10$^{25}$~W~Hz$^{-1}$. The 3--6~GHz spectral index of this source is steep (or ultra-steep by some definitions, $\alpha$~=~$-$1.13~$\pm$~0.05). This is a characteristic known to be an effective method to search for powerful radio sources at high redshift, although the physical explanation is not conclusively understood \citep[e.g.][]{singh14}. The source is deemed to be resolved, although it is well characterised by a single Gaussian component, and there is no evidence for large-scale jet emission in either the S-HIGH or S-LOW images. This source is thus likely a newly-discovered powerful high-$z$ radio galaxy, and worthy of follow-up observations.

\subsection{Intrinsic radio sizes}

\begin{figure}[h]
\centering
\includegraphics[width=\columnwidth]{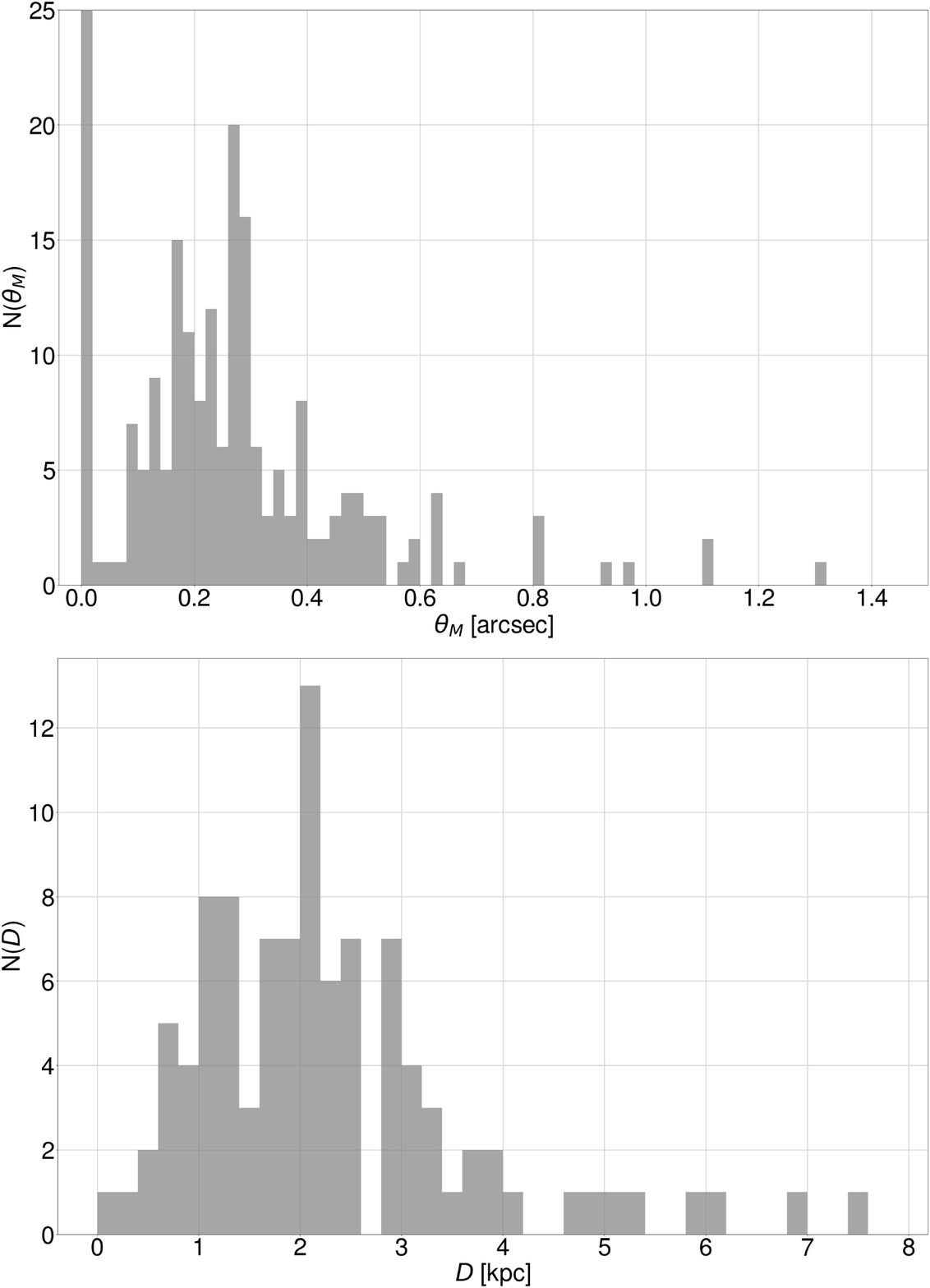}
\caption{Distribution of the FWHM deconvolved source major axes as measured from the C-HIGH images (upper panel). The first bin (zero) contains 83 sources for which reliable deconvolved sizes could not be obtained. Not including these sources, the median deconvolved source size is 0$''$.27, with a median absolute deviation of 0$''$.1. The lower panel shows the distribution for physical source sizes.}
\label{fig:csizes}
\end{figure}

The C-HIGH images with angular resolutions of $\sim$0.3--0$''$.5 will offer the best means for estimating intrinsic source sizes from this project. To examine this we present the distribution of the FWHM deconvolved source major axes in the upper panel of Figure \ref{fig:csizes}. The median deconvolved source size ($\theta_{M}$) (for entries in the catalog that have non-zero values) is 0$''$.27~$\pm$~0$''$.25. For the assumed cosmology, the radio components with associated redshift measurements have a median physical size of 1.9~$\pm$~1.2 kpc. The corresponding distribution in physical units is shown in the lower panel of Figure \ref{fig:csizes}.

Higher angular resolution observations reaching $\sim$0.6 $\mu$Jy beam$^{-1}$ have been made of the GOODS North field at X-band \citep[10 GHz;][]{murphy17}. This study also used long-track A- and C-configuration VLA observations, resulting in a measurement of $\langle\theta_{M}\rangle$~=~0$''$.167~$\pm$~0$''$.032 (with RMS scatter of 0$''$.91), corresponding to a median linear size of 1.3~$\pm$~0.28 kpc (with RMS scatter 0.79~kpc). This was shown to match both the sizes of the dust emission regions measured at sub-mm wavelengths, and the sizes measured in extinction-corrected H$\alpha$ imaging. 

Although consistent with the 1$\sigma$ measurement errors, and despite the high angular resolution of the C-HIGH images, our median linear size is larger than that measured by \citet{murphy17}. This suggests that even half-arcsecond resolution is sub-optimal for robust size measurements of high-redshift star-forming galaxies at radio wavelengths, although a larger angular size is expected to be seen at lower radio frequencies due to cosmic-ray propagation effects. For a more detailed analysis of radio source sizes using our data, we again refer the reader to Jim\'{e}nez-Andrade et al. (\emph{submitted}). The interferometers will not resolve out significant emission on these scales due to insufficient short spacings. Neglecting AGN contamination, the radio emission at both 6~GHz and 10~GHz should be arising from the star-forming regions. Given that the angular resolution of both of these studies exceeds that of Band 2 of SKA-MID (0.95 -- 1.76 GHz), X-band studies with the A-configuration of the VLA are likely the best path to such intrinsic size measurements prior to the arrival of the Next-Generation VLA \citep[ngVLA;][]{murphy18,mckinnon19}.

\subsection{Optical colours of the hosts of compact radio sources}

\begin{figure*}[ht!]
\includegraphics[width=\textwidth]{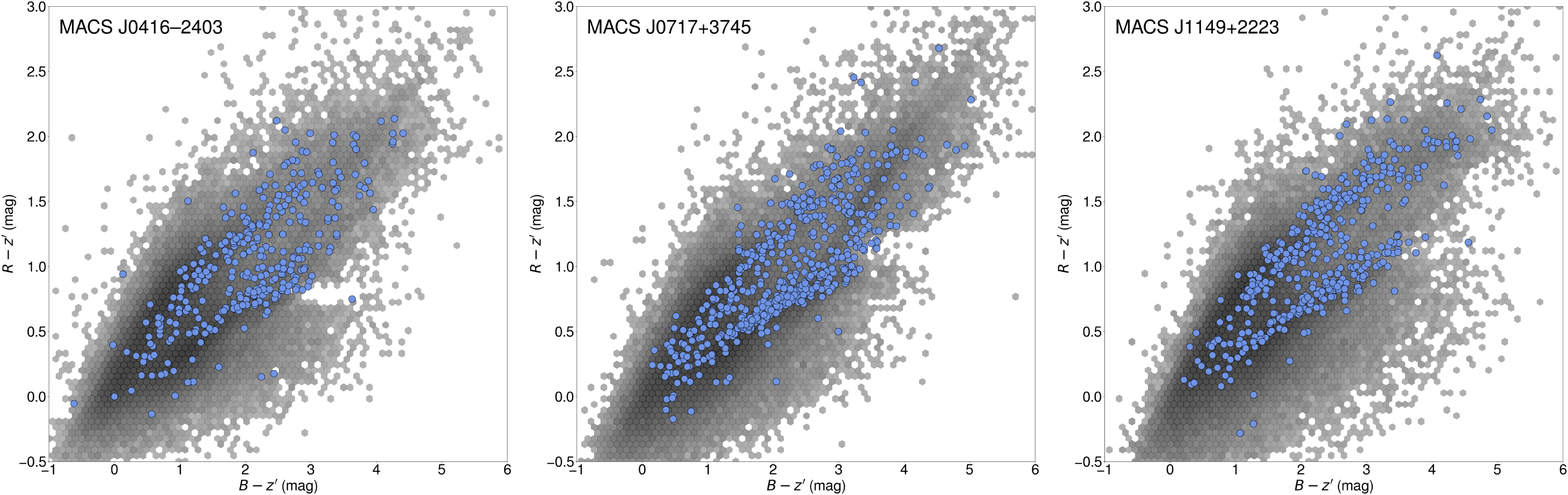}
\caption{Apparent $B-z'$ against $R-z'$ magnitudes from the Subaru catalogs for the three Frontier Field clusters. The colours of the galaxies hosting compact radio components are shown with the blue markers. All other entries in the Subaru photometric catalogs are shown in the background 2D histogram in grey. The radio detections are representative of the general population of galaxies detected in the deep optical / near-infrared imaging.}
\label{fig:opticalcolours}
\end{figure*}

Figure \ref{fig:opticalcolours} shows the apparent magnitude differences of bands $B-z'$ against $R-z'$ from the Subaru catalogs. The hosts of compact radio sources are shown in blue. The rest of the entries in the Subaru catalogs are as a grey 2D histogram. The use of apparent instead of intrinsic magnitudes causes the significant scatter in the latter distribution, however red (upper) and blue (lower) branches are evident. With reference also to Figure \ref{fig:z_dist}, the compact radio catalog is mostly detecting galaxies with redshifts higher than those of the clusters (although not necessarily on sight lines that pass through the mass distribution of the cluster). Despite the use of apparent magnitudes, the colour distributions of the galaxies with radio detections are also visible. The radio data are deep enough to detect large numbers of galaxies in the blue cloud, i.e.~regular star-forming spiral galaxies.

\subsection{The radio relic in MACSJ1149.5+2223}
\label{sec:relics}

\begin{figure*}[ht!]
\includegraphics[width=\textwidth]{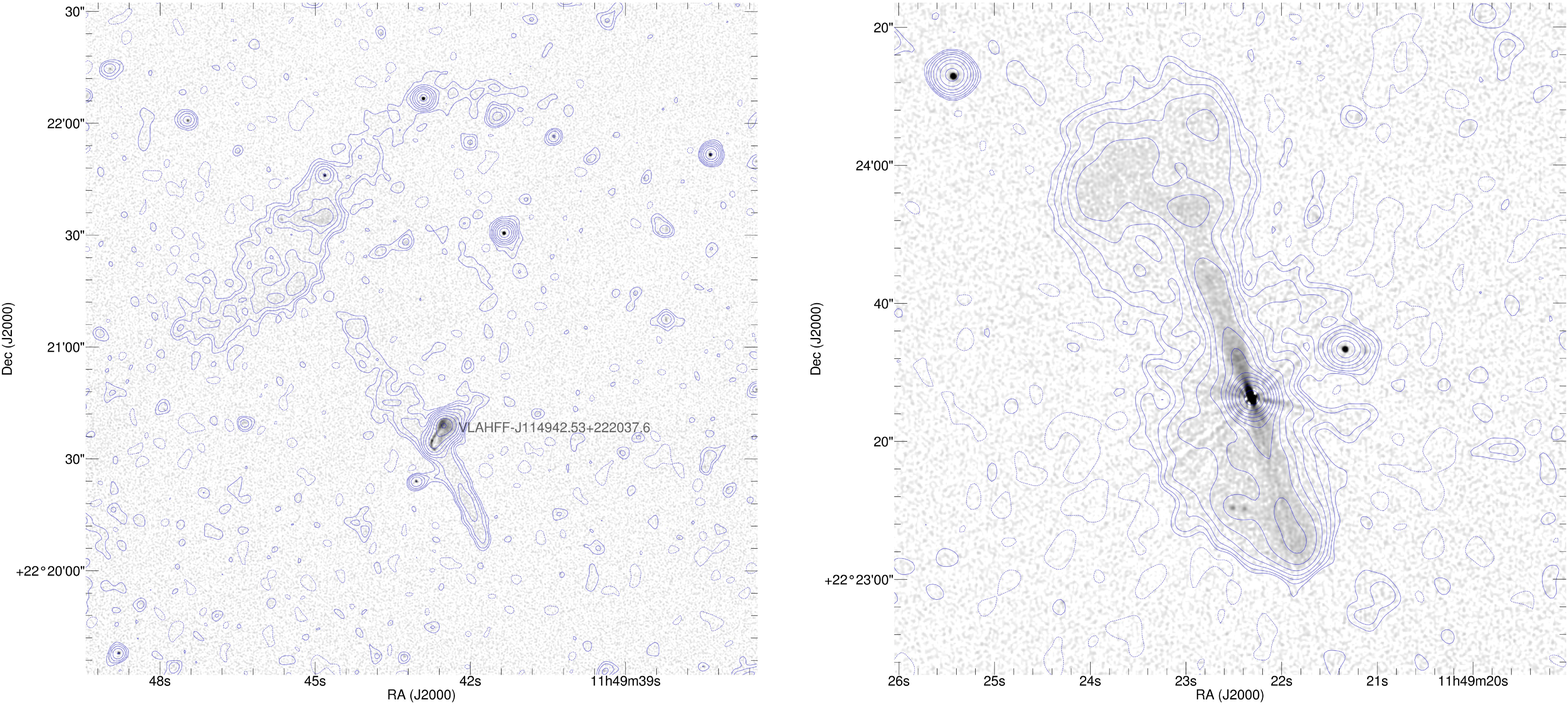}
\caption{Diffuse radio structures on the periphery of MACSJ1149. Both panels feature greyscale images of the S-HIGH maps. The greyscale runs from 0.0 (white) to 40 $\mu$Jy beam$^{-1}$ black, and has a square-root stretch function. The contours trace the S-LOW image with 3$''$ resolution. Contour values are 4 $\mu$Jy beam$^{-1}$ $\times$ 3$^{i/2}$, for $i = \{0,1,2,3,4\ldots\}$. The left hand panel shows the eastern radio relic. Our imaging reveals that the narrow angle tail source VLAHFF-J114942.53+222037.6 is embedded in this feature. The right hand panel shows the radio source VLAHFF-J114922.34+222327.7. This was previously thought to be a second relic in MACSJ1149, but is revealed to be a FR-I source at a lower redshift. \label{fig:relics}}
\end{figure*}

The diffuse radio emission associated with galaxy clusters is generally observed at low radio frequencies due to the low surface brightness and steep radio spectra, however our 3~GHz observations are deep enough to provide images of these structures at high angular resolution.

The central right-hand panel of Figure \ref{fig:maps}, and Figure \ref{fig:vlahst}, are powerful illustrations of the usefulness of high angular resolution for relic and halo studies. MACSJ0717 hosts one of the brightest radio haloes known \citep[e.g.][]{vanweeren17,bonafede18}, and although the S-HIGH image resolves out the largest scales of the diffuse halo, the filamentary structures in the chair-shaped relic surrounding the central narrow-angle tailed radio galaxy can be seen with detail not possible with current low frequency instruments. The central tailed radio source has a host galaxy with a redshift that places it within the cluster, however \citet{rajpurohit20} use spectral modelling to suggest that it is not interacting with the relic itself, and is merely seen in projection.

\citet{bonafede12} present observations of MACSJ1149 at 323 MHz and 1.4 GHz, concluding the cluster hosts a double radio relic. Double relic structures are relatively rare, thought to arise when an on-going cluster merger is seen from a favourable viewpoint \citep{bonafede17}. Figure \ref{fig:relics} shows the view of the relevant regions afforded by our 3~GHz data. The greyscale is the S-HIGH image, overlaid with the contours of the 3$''$ resolution S-LOW image (please refer to the figure caption for further details). The left-hand panel shows the eastern relic, which is revealed to be a complex structure that is resolved into two distinct, almost perpendicular components. The bright peak seen at lower frequencies is actually an embedded NAT source (VLAHFF-J114922.34+222327.7, panel number 60 on Figure \ref{fig:extended2}). At $z$~=~0.545 this source could truly be embedded in the relic, rather than being a foreground or background source seen in projection. The right hand panel shows the region of the reported second relic in MACSJ1149, which in our high-resolution observations is revealed to be a FR-I source, VLAHFF-J114922.34+222327.7 (panel number 53 on Figure \ref{fig:extended2}). The high-resolution radio data also allows us to provide an unambiguous optical identification for the host galaxy, which has a photometric redshift of 0.24, and so this entire radio structure is not actually associated with the cluster.

\section{Conclusions}
\label{sec:conclusions}

Using the VLA at 3 and 6 GHz, we have made some of the deepest ($\sim$1 $\mu$Jy beam$^{-1}$), high-resolution ($\sim$0$''$.5) radio images to date of the three MACS strong gravitational lensing clusters from the Frontier Fields program: MACSJ0416.1$-$2403, MACSJ0717.5+3745, and MACSJ1149.5+2223. From these images we have derived catalogs with a total of 1,966 compact radio components, 1,296 of which have identified optical hosts. Relevant properties from the optical / near-infrared data have been collated with the radio properties into a unified radio and optical / near-infrared catalog. We make use of the most recent mass and shear models available for the Frontier Fields to estimate the gravitational lensing magnification of the radio components with redshifts beyond those of the clusters. From this analysis we detect a total of 13 moderately lensed (2.1~$<$~$\mu$~$<$~6.5) sources. The optical / near-infrared colours of the radio detections show that we are detecting a significant population of regular, blue star-forming galaxies at high redshift.

The dual-frequency radio observations provide cataloged 3--6 GHz spectral index measurements for 169 compact components within the 50\% level of the C-band primary beam. These are mostly dominated by sources with typical synchrotron spectra, although there is a significant population of flat and inverted spectrum sources. We also provide a catalog of intrinsic source sizes measured from the highest resolution ($\sim$0$''$.3) C-band images. Our median angular size of 0$''$.27~$\pm$~0$''$.25 is somewhat larger than the 0$''$.167~$\pm$~0$''$.032 value measured at higher resolution by \citet{murphy17}, and while our measurements offer good constraints such work is likely better pursued using higher frequency observations with the VLA's most extended configurations.

A total of 66 radio sources with extended morphologies were also identified, which are a mixture of sources hosting radio jets of numerous types, resolved spiral disks, and circumnuclear star formation. The advantage of using high angular resolution imaging for studies of the diffuse emission associated with galaxy clusters is exemplified by the detailed imaging of the diffuse emission associated with MACSJ0717 and MACSJ1149. In the case of the latter we reveal that a bright feature in the radio relic previously observed a low frequencies is actually an embedded narrow angle tail source, and the that putative second relic in this cluster is actually a FR-I type radio galaxy at an intervening redshift.

The VLA Frontier Fields survey is a public legacy project, and we make all our catalog and image products freely available at:\\ 
\url{https://science.nrao.edu/science/surveys/vla-ff}.\\

\acknowledgments

We thank the anonymous referees for taking the time to read and comment on this paper. The National Radio Astronomy Observatory is a facility of the National Science Foundation operated under cooperative agreement by Associated Universities, Inc. Based on observations made with the NASA/ESA Hubble Space Telescope, obtained from the data archive at the Space Telescope Science Institute. STScI is operated by the Association of Universities for Research in Astronomy, Inc. under NASA contract NAS 5-26555. Support for this work was provided by NASA through grant number HST-AR-14306.001-A from the Space Telescope Science Institute, which is operated by AURA, Inc., under NASA contract NAS 5-26555. This work utilizes gravitational lensing models produced by PIs Bradač, Natarajan \& Kneib (CATS), Merten \& Zitrin, Sharon, Williams, Keeton, Bernstein and Diego, and the GLAFIC group. This lens modeling was partially funded by the \emph{HST} Frontier Fields program conducted by STScI. STScI is operated by the Association of Universities for Research in Astronomy, Inc. under NASA contract NAS 5-26555. The lens models were obtained from the Mikulski Archive for Space Telescopes (MAST). This work is based in part on data collected at Subaru Telescope, which is operated by the National Astronomical Observatory of Japan. IH acknowledges support from the UK Science and Technology Facilities Council [ST/N000919/1]; the Oxford Hintze Centre for Astrophysical Surveys which is funded through generous support from the Hintze Family Charitable Foundation; and the South African Radio Astronomy Observatory which is a facility of the National Research Foundation (NRF), an agency of the Department of Science and Innovation.

\software{Astropy \citep{astropy:2013, astropy:2018}, CASA \citep{mcmullin07}, Matplotlib \citep{hunter07}, ProFound \citep{robotham18}, PyBDSF \citep{mohan15}, wsclean \citep{offringa14}}

%

\vspace{5mm}
\facilities{Karl G.~Jansky Very Large Array, \emph{Hubble} Space Telescope, Subaru Telescope}





\appendix

\section{Checking the optical vs radio astrometric frames}
\label{sec:astrometry}

Positions of compact radio components that are determined by the fitting of a Gaussian (or point) component to the pixel brightness distribution are subject to statistical uncertainties that are related to the resolution of the instrument as well as the signal to noise ratio (SNR) of the detection \citep{condon97}. In contrast, high resolution optical images may contain morphologies that may well be more complex than the radio. In this case, automated source extraction tends to emphasize the fitting of apertures that are appropriate for (often multi-band) photometric measurements. The position of the galaxy then tends to be determined by using a brightness-weighted mean of the pixels within the aperture. Thus, even if the radio and optical emission has a co-spatial origin, the automatically-measured positions of the same object may differ slightly between the two wavelengths due to both the differing source extraction methods, as well as the statistical jitter that influences both with decreasing SNR (as already introduced via Equation \ref{eq:pos_scatter}). Additionally, radio and optical instruments may also have different absolute astrometric reference frames due to a variety of factors.

\begin{figure}[h]
\centering
\includegraphics[width=\columnwidth]{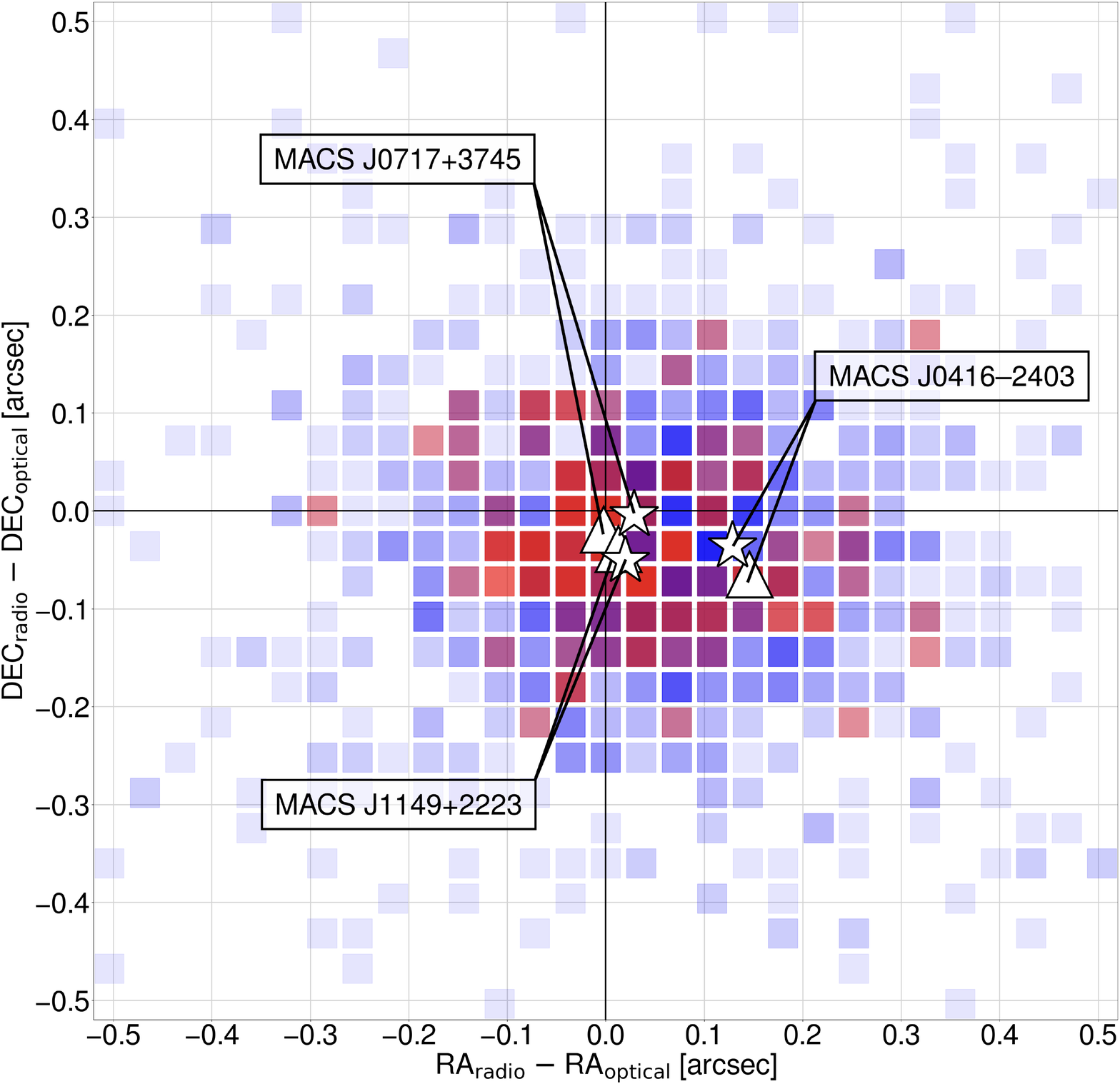}
\caption{Differences between the radio and optical positions in right ascension and declination, for both the DeepSpace (red) and CLASH (blue) catalogs. The mean values of the offset distributions for the three clusters (as labelled) are also shown for both the DeepSpace (triangles) and CLASH (stars) catalogs. Note that the mean values are in all cases significantly smaller than the resolution element of the S-HIGH image which was used as the starting point for generating the catalogs, as described in Section \ref{sec:results}. The resolution of the S-HIGH images ($\sim$0$''$.5 -- 1$''$ depending on the cluster declination) is between half of the extent of the above figure and its full extent.}
\label{fig:astrometry}
\end{figure}

Figure \ref{fig:astrometry} shows the checks we have performed to demonstrate that the above factors are not significant enough to influence the reliability of the cross-matching process, i.e. that the offsets in both an absolute and statistical sense are significantly smaller than the resolution element of the radio observations.

The difference between the radio and optical positions in both right ascension and declination is determined for both the DeepSpace (red) and CLASH (blue) catalogs and plotted as a 2D histogram. In addition to this, the mean values per cluster and per catalog are also plotted. In all cases the mean offsets are smaller than the resolution element of the S-HIGH images used to extract the catalog.


\section{Deconvolved source sizes and their uncertainties when observed with an elliptical beam}
\label{sec:sizes}

In the case of a circular restoring beam, the deconvolved full-width half-maximum (FWHM) of a source is given by
\begin{equation}
\theta_{\mathrm{M}}~=~(\phi^{2} - \theta^{2}_{\mathrm{beam}})^{1/2},
\label{eq:phi_circ}
\end{equation}
\noindent
where $\phi$ is the FWHM of the fitted component in the image, and $\theta_{\mathrm{beam}}$ is the FWHM of the restoring beam. The error of $\theta_{\mathrm{M}}$ is therefore
\begin{equation}
\sigma_{\theta_{\mathrm{M}}}~=~\sigma_{\phi}\left[1~-~\left(\frac{\theta_{\mathrm{beam}}}{\phi}\right)^{2}\right]^{-1/2}.
\label{eq:sigma_phi_circ}
\end{equation}

\begin{figure}[h]
\centering
\includegraphics[width=\columnwidth]{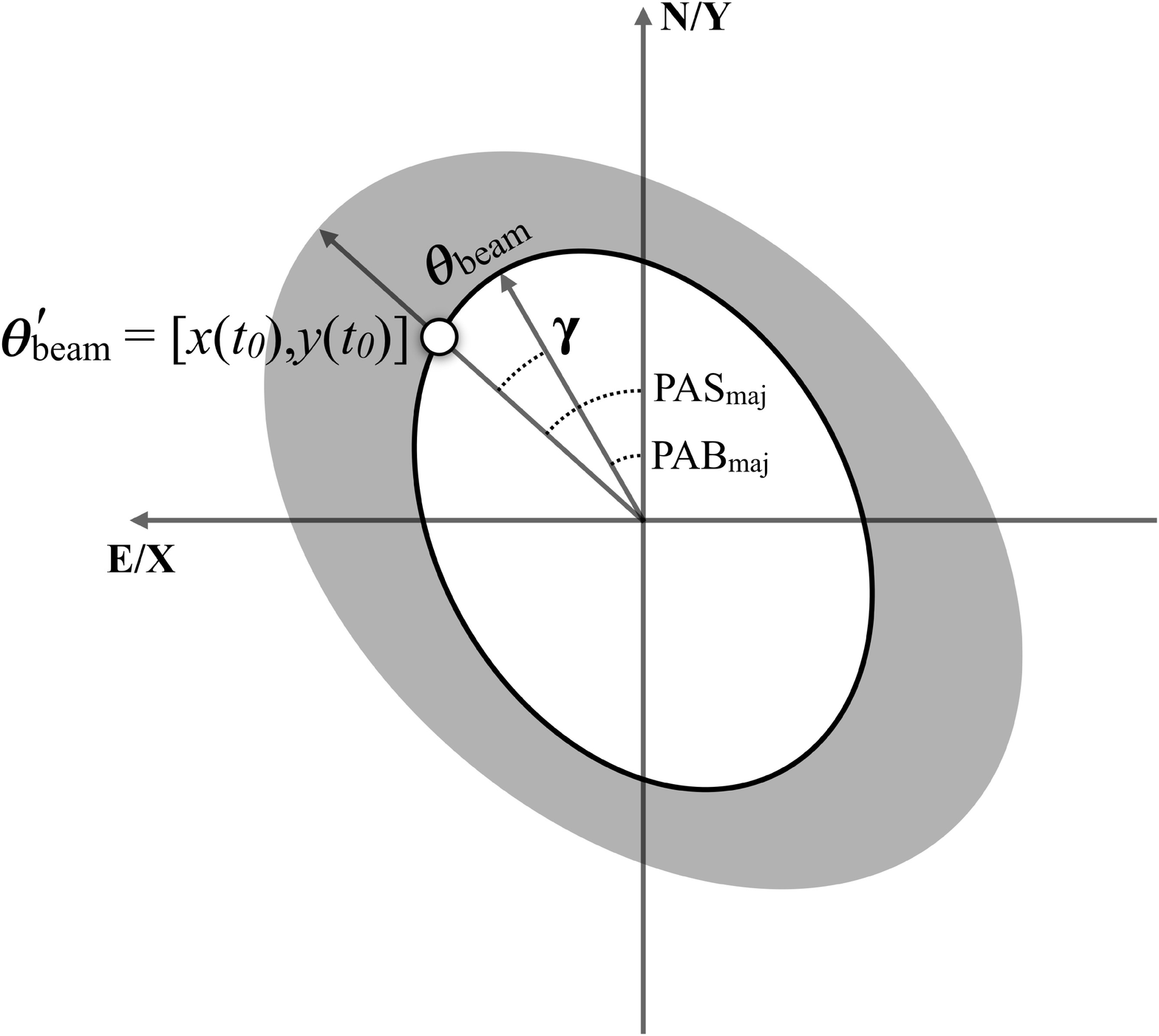}
\caption{A 2D Gaussian radio source (the FWHM of which is represented by the grey ellipse) that has been observed with a 2D Gaussian restoring beam (the FWHM of which is the white ellipse). The arrows show the position angle of the FWHM major axis of the beam (PAB$_{\mathrm{maj}}$) and the source (PAS$_{\mathrm{maj}}$). Equations \ref{eq:ell_x} and \ref{eq:ell_y} can be evaluated at $t~=~t_{0}$ in order to infer the length of the beam along the direction of the source major axis ($\theta^{'}_{\mathrm{beam}}$) via Equation \ref{eq:tgamma}.}
\label{fig:ellipses}
\end{figure}

In the case of an elliptical beam, estimating $\theta_{\mathrm{M}}$ and $\sigma_{\theta_{\mathrm{M}}}$ along the major axis of the source can be done with Equations \ref{eq:phi_circ} and \ref{eq:sigma_phi_circ} but only if the position angle of the restoring beam matches that of the source. For determining the deconvolved size ($\theta^{'}_{\mathrm{M}}$) of a source along its major axis and the associated uncertainty ($\sigma^{'}_{\theta_{\mathrm{M}}}$) in the general case, we have to determine the FWHM of the elliptical beam projected along the major axis of the source. This vector will be at a position angle $\theta^{'}_{\mathrm{beam}}$, see Figure \ref{fig:ellipses}. The parametric equations of an ellipse are
\begin{eqnarray}
x(t)&~=~&a~\mathrm{cos}(t)\mathrm{cos}\left[\mathrm{PAS_{maj}}(\pi/180) - (\pi/2)\right] \nonumber \\
    & & - b~\mathrm{sin}(t)\mathrm{sin}\left[\mathrm{PAS_{maj}}(\pi/180) - (\pi/2)\right]
\label{eq:ell_x}
\end{eqnarray}
and
\begin{eqnarray}
y(t)&~=~&a~\mathrm{cos}(t)\mathrm{sin}\left[\mathrm{PAS_{maj}}(\pi/180) - (\pi/2)\right] \nonumber \\
    & & + b~\mathrm{sin}(t)\mathrm{cos}\left[\mathrm{PAS_{maj}}(\pi/180) - (\pi/2)\right]
\label{eq:ell_y}
\end{eqnarray}
where $0~\leq~t~\leq~2\pi$. If $a$ and $b$ are the FWHM major and minor axes of the source then
\begin{equation}
t~=~\mathrm{arctan}\left[\left(\frac{a}{b}\right)\mathrm{tan}(\gamma)\right],
\label{eq:tgamma}
\end{equation}
where $\gamma$ is the angle between the major axis of the ellipse and the vector to the point on the ellipse at $[x(t),y(t)]$. The vector on the beam ellipse that is aligned with the source major axis is therefore determined by evaluating Equations \ref{eq:ell_x}, \ref{eq:ell_y} and \ref{eq:tgamma} for $\gamma~=~\mathrm{PAB}_{\mathrm{maj}} - \mathrm{PAS}_{\mathrm{maj}}$, where all angles are given in degrees, and the standard convention of measuring position angles east-of-north applies. Once $x(t_{0})$ and $y(t_{0})$ are known the FWHM of the elliptical beam along the source major axis is simply
\begin{equation}
\theta^{'}_{\mathrm{beam}}~=~\left[x(t_{0})^{2} + y(t_{0})^{2}\right]^{1/2}
\end{equation}
and its associated uncertainty is approximated as
\begin{equation}
\sigma^{'}_{\theta_{\mathrm{M}}}~=~\sigma_{\phi^{'}}\left[1 - \left(\frac{\theta^{'}_{\mathrm{beam}}}{\phi^{'}}\right)^{2}\right]^{-1/2},
\end{equation}
where $\phi^{'}$ and $\sigma_{\phi^{'}}$ are the source major axis at FWHM and its associated uncertainty, as measured by {\tt PyBDSF}.

\section{Structure of the compact component catalogs}
\label{sec:radoptstructure}

MACSJ0416, MACSJ0717 and MACSJ1149 contain, respectively, 343, 535 and 418 radio sources that are well-described by a point or Gaussian component and have associated optical identifications. These 1,296 components each have entries in a per-cluster table, the columns of which are described below. Each cluster also has a table of radio components that do not have \emph{cataloged} optical identifications within 1$''$. For MACSJ0416, MACSJ0717 and MACSJ1149, the total counts of radio components of this type are 162, 328 and 182 correspondingly. These tables share columns 0--33 inclusive of the optically-matched catalog.\\
\\
\noindent
{\bf (0)}: A unique positional identifier for each component of the form VLAHFF-JHHMMSS.SS+/-DDMMSS.S.
\\

\noindent
{\bf (1)}: The J2000 Right Ascension of the component in decimal degrees.
\\

\noindent
{\bf (2)}: The J2000 Declination of the component in decimal degrees.
\\

\noindent
{\bf (3) and (4)}: The 1$\sigma$ uncertainties in the position of the component in arcseconds. Note that these are uncertainties derived from the component fitting routines of {\tt PyBDSF}, and as such do not include any systematic astrometric offsets.
\\

\noindent
{\bf (5) and (6)}: Integrated S-band flux density and 1$\sigma$ uncertainty in $\mu$Jy.
\\

\noindent
{\bf (7) and (8)}: Peak S-band brightness and 1$\sigma$ uncertainty in $\mu$Jy~beam$^{-1}$.
\\

\noindent
{\bf (9) and (10)}: `Best' S-band flux density (S$^{*}$) and 1$\sigma$ uncertainty in $\mu$Jy~beam$^{-1}$. Please refer to Section \ref{sec:s-star} for further details.
\\

\noindent
{\bf (11)}: Assumed S-band primary beam correction factor that has been applied to columns (5) to (10) inclusive.
\\

\noindent
{\bf (12) and (13)}: Integrated C-band flux density and 1$\sigma$ uncertainty in $\mu$Jy.
\\

\noindent
{\bf (14) and (15)}: Peak C-band brightness and 1$\sigma$ uncertainty in $\mu$Jy~beam$^{-1}$.
\\

\noindent
{\bf (16) and (17)}: `Best' C-band flux density and 1$\sigma$ uncertainty in $\mu$Jy~beam$^{-1}$.
\\

\noindent
{\bf (18)}: Assumed C-band primary beam correction factor that has been applied to columns (12) to (17) inclusive.
\\

\noindent
{\bf (19) and (20)}: Spectral index ($\alpha$) and 1$\sigma$ uncertainty for the component, derived from the S-band and matched-resolution C-band catalogs. Please refer to Section \ref{sec:alphas} for details.
\\

\noindent
{\bf (21), (22), (23), (24), (25) and (26)}: Major and minor axes (in arcseconds) and position angle (in degrees, east of north) of the fitted components in the S-HIGH images, together with the associated 1$\sigma$ uncertainties, as derived from the component fitting routines in PyBDSF. Values of 0.0 in these columns mean the component is unresolved in one or more dimensions.
\\

\noindent
{\bf (27) and (28)}: Major axis and associated uncertainty of the deconvolved components in the S-HIGH images. The major axis is the source-finder's estimates of the intrinsic angular extent of the radio component, however the uncertainties in the deconvolved size are determined using the method described in Section \ref{sec:s-star}. As above, values of 0.0 in these columns mean the component is unresolved in one or more dimensions.
\\

\noindent
{\bf (29)}: A Boolean flag which is 1 if the source is deemed to be reliably resolved, 0 otherwise, based on the projection of the source major axis along the major axis of the restoring beam. See Section \ref{sec:s-star} for further details.
\\

\noindent
{\bf (30)}: ID of the assumed optical host galaxy in the DeepSpace \citep{shipley18} catalogs. Note that in the DeepSpace catalogs these IDs begin from zero for each cluster, and so are not unique across all fields.
\\

\noindent
{\bf (31) and (32)}: Right Ascension and Declination in decimal degrees of the assumed optical host galaxy in the DeepSpace catalogs.
\\

\noindent
{\bf (33)}: Radial separation of the DeepSpace match in arcseconds.
\\

\noindent
{\bf (34)}: Spectroscopic redshift of the DeepSpace match, where available.
\\

\noindent
{\bf (35), (36) and (37)}: Preferred photometric redshift and its lower and upper 68\% confidence intervals of the DeepSpace match as returned by the EAZY code \citep{brammer08}, where available.
\\

\noindent
{\bf (38):} Log stellar mass of the host galaxy (M$_{\odot}$), as listed in the DeepSpace catalogs, derived using the FAST code \citep{kriek18}.
\\

\noindent
{\bf (39):} Log star-formation rate of the host galaxy (M$_{\odot}$ year$^{-1}$), as listed in the DeepSpace catalogs, derived using the FAST code \citep{kriek18}.
\\

\noindent
{\bf (40):} Log specific star-formation rate of the host galaxy (M$_{\odot}$ year$^{-1}$), as listed in the DeepSpace catalogs, derived using the FAST code \citep{kriek18}.
\\

\noindent
{\bf (41)}: ID of the assumed optical host galaxy in the CLASH catalogs.
\\

\noindent
{\bf (42) and (43)}: Right Ascension and Declination in decimal degrees of the assumed optical host galaxy in the CLASH catalogs.
\\

\noindent
{\bf (44)}: Radial separation of the CLASH match in arcseconds.
\\

\noindent
{\bf (45) and (46)}: CLASH Subaru $z$-band magnitude of the assumed host, and uncertainty.
\\

\noindent
{\bf (47), (48) and (49)}: Bayesian Photo-$z$ (BPZ) redshift computed from the Subaru photometry, as well as the upper and lower limits. Please refer to Umetsu et al.~(2014) and Medezinski et al.~(2013) for details about the CLASH data, and the determination of photometric redshifts.
\\

\noindent
{\bf (50)}: Median lensing magnification value derived from the best available redshift. Preference is given to spectroscopic redshifts, however the photometric values are used for the majority of objects where only such measurements are available. 
\\

\noindent
{\bf (51)}: Median absolute deviation of the magnification values used to calculate column (51).
\\

\noindent
{\bf (52)}: The number of independent weak lensing shear and mass surface density models used to compute the values in columns (50) and (51) inclusive. Values derived from several models tend to be closer to the high magnification regions in the center of the cluster, covering the \emph{HST} area. Please refer to Section \ref{sec:mags} for details of the magnification calculations.

\section{Radio sources with extended morphologies}
\label{sec:extendedappendix}

\subsection{Optical and radio overlays}

The three sections of Figure \ref{fig:extended1} show the S-HIGH contours for the 66 sources identified as having complex radio morphologies, overlaid on a RGB image formed from the Subaru $B$, $R$, and $z$ band images. Please refer to the caption of Figure \ref{fig:extended3} for further details. Note that these images are not primary beam corrected, in the interests of achieving a uniform contouring scheme, and are presented primarily for morphological classification. The properties of each of the complex sources are provided in order of decreasing total (primary-beam corrected) flux density in Tables \ref{tab:extended0416}, \ref{tab:extended0717} and \ref{tab:extended1149} for MACSJ0416, MACSJ0717 and MACSJ1149 respectively. Here we provide brief comments on the radio morphology and optical host.\\

\subsection{Properties of extended radio sources}

The properties of radio sources with extended morphologies are presented in Tables \ref{tab:extended0416}, \ref{tab:extended0717} and \ref{tab:extended1149} for MACSJ0416, MACSJ0717 and MACSJ1149 respectively. The equatorial (J2000) positions are the flux-weighted centroids of the emission associated with each extended source, as determined by {\tt ProFound}. Optical IDs from the CLASH catalogs together with any available photometric redshifts are also listed, together with the integrated flux densities (and associated uncertainty) as measured from the S-HIGH and S-LOW images. The latter provides a more robust estimate of the total integrated flux density of these sources at 3~GHz. As can be seen from the tables, the extended A-configuration of the VLA (coupled with the weighting scheme required to deliver the high angular resolution required for the primary goal of these observations) resolves out a significant amount of extended emission.

\begin{table*}
\begin{minipage}{175mm}
\centering
\caption{Positions and integrated flux densities of the extended radio sources in MACSJ0416. Unique identifiers with the same format as described in Appendix \ref{sec:radoptstructure} are based on the flux-weighted centroids of the radio emission as identified by the {\tt ProFound} source finder. Optical host IDs from the CLASH data are provided, along with photometric redshift estimates where available. Integrated flux density measurements (and associated uncertainties) are provided from both the S-HIGH and S-LOW images, with the latter providing a more robust estimate of the total integrated flux density at 3~GHz.}
\begin{tabular}{lllllllll} \hline \hline
ID & RA & DEC & CLASH ID & $z_{\mathrm{BPZ}}$ & S$_{int}^{\mathrm{HIGH}}$ & $\sigma$S$_{int}^{\mathrm{HIGH}}$& S$_{int}^{\mathrm{LOW}}$ & $\sigma$S$_{int}^{\mathrm{LOW}}$ \\
 & [deg] & [deg] & &  & [mJy] & [mJy]& [mJy]& [mJy] \\ \hline
VLAHFF-J041514.28-240934.3 & 63.80952 & $-$24.15953 & -    & -     & 2.6 & 0.05 & - & - \\
VLAHFF-J041528.24-240913.7 & 63.86767 & $-$24.15382 & -      & -    & 3.14 & 0.07 & 17.26 & 0.17 \\
VLAHFF-J041549.94-235437.3 & 63.95812 & $-$23.91036 & -     & -     & 78.62 & 0.09 & 318.61 & 0.27 \\
VLAHFF-J041553.04-240716.5 & 63.97102 & $-$24.12125 & 43053 & 0.499 & 14.09 & 0.12 & 86.8 & 0.13 \\
VLAHFF-J041554.13-240406.9 & 63.97557 & $-$24.0686 & 53515 & 1.645 & 33.52 & 0.03 & 103.55 & 0.05 \\
VLAHFF-J041556.77-240434.5 & 63.98655 & $-$24.07626 & 54833 & 1.695 & 1.2 & 0.03 & 16.9 & 0.08 \\
VLAHFF-J041603.56-240429.2 & 64.01487 & $-$24.07478 & 51115 & 0.945 & 2.73 & 0.02 & 10.46 & 0.04 \\
VLAHFF-J041603.63-240551.4 & 64.01517 & $-$24.09761 & 48556 & 0.392 & 6.80 & 0.04 & 20.23 & 0.05 \\
VLAHFF-J041604.63-240415.0 & 64.01931 & $-$24.07086 & 53930 & 0.037 & 4.76 & 0.03 & 13.78 & 0.04 \\
VLAHFF-J041604.84-241028.0 & 64.0202 & $-$24.17446 & 32639 & 0.288 & 16.91 & 0.02 & 53.73 & 0.06 \\
VLAHFF-J041605.95-235733.6 & 64.0248 & $-$23.95933 & 77422 & 0.273 & 2.05 & 0.06 & 4.56 & 0.06 \\
VLAHFF-J041609.06-240945.0 & 64.03779 & $-$24.16253 & 33689 & 0.025 & 2.07 & 0.03 & 6.58 & 0.06 \\
VLAHFF-J041609.16-240402.8 & 64.03821 & $-$24.06745 & 56589 & 0.438 & 12.62 & 0.02 & 40.61 & 0.05 \\
VLAHFF-J041609.64-240555.3 & 64.04018 & $-$24.09871 & 47184 & 0.945 & 2.86 & 0.04 & 7.08 & 0.04 \\
VLAHFF-J041613.94-235645.0 & 64.05811 & $-$23.94585 & 81839 & 0.024 & 10.15 & 0.08 & 29.03 & 0.09 \\
VLAHFF-J041620.05-241008.0 & 64.08358 & $-$24.1689 & 31008 & 1.176 & 112.72 & 0.08 & 372.75 & 0.18 \\
VLAHFF-J041621.02-235804.8 & 64.0876 & $-$23.96803 & 75399 & 1.438 & 12.14 & 0.04 & 41.48 & 0.07 \\
VLAHFF-J041625.22-240011.6 & 64.10509 & $-$24.00324 & 68394 & 0.516 & 37.16 & 0.025 & 118.81 & 0.07 \\
VLAHFF-J041628.34-235321.9 & 64.11809 & $-$23.88944 & 99186 & 0.358 & 11.14 & 0.26 & 57.44 & 0.55 \\
VLAHFF-J041629.22-235903.1 & 64.12177 & $-$23.9842 & 72655 & 0.574 & 2.6 & 0.05 & 6.48 & 0.06 \\
VLAHFF-J041632.03-240340.8 & 64.13346 & $-$24.06134 & 55131 & 0.874 & 2.26 & 0.03 & 7.54 & 0.06 \\
VLAHFF-J041633.46-240857.7 & 64.13944 & $-$24.14937 & 35525 & 1.184 & 9.85 & 0.03 & 32.74 & 0.08 \\
VLAHFF-J041646.28-235623.4 & 64.19284 & $-$23.93984 & 81280 & 3.53 & 164.61 & 1.1 & 272.8 & 0.42 \\
VLAHFF-J041648.84-240243.4 & 64.20352 & $-$24.04541 & 58828 & 0.831 & 131.98 & 0.08 & 482.07 & 0.3 \\
VLAHFF-J041649.20-235802.6 & 64.20504 & $-$23.9674 & 77823 & 0.423 & 41.31 & 0.28 & 141.28 & 0.29 \\
VLAHFF-J041656.65-240600.4 & 64.23607 & $-$24.10013 & 46835 & 1.148 & 295.60 & 0.37 & 1109.11 & 0.78 \\ \hline
\end{tabular}
\label{tab:extended0416}
\end{minipage}
\end{table*}

\begin{table*}
\begin{minipage}{175mm}
\centering
\caption{Positions and integrated flux densities of the extended radio sources in MACSJ0717. Please refer to the caption of Table \ref{tab:extended0416} for further details.}
\begin{tabular}{lllllllll} \hline \hline
ID & RA & DEC & CLASH ID & $z_{\mathrm{BPZ}}$ & S$_{int}^{\mathrm{HIGH}}$ & $\sigma$S$_{int}^{\mathrm{HIGH}}$& S$_{int}^{\mathrm{LOW}}$ & $\sigma$S$_{int}^{\mathrm{LOW}}$ \\
 & [deg] & [deg] & &  & [mJy] & [mJy]& [mJy]& [mJy] \\ \hline
VLAHFF-J071632.28+373912.5 & 109.1345 & 37.65348 & 26801 & 2.813   & 301.87 & 0.34 & 396.62 & 0.47 \\
VLAHFF-J071632.65+374251.6 & 109.13604 & 37.71434 & -     & -      & 25.63  & 0.14 & 89.57 & 0.19 \\
VLAHFF-J071644.28+373956.1 & 109.18451 & 37.66561 & 43871 & 0.076  & 74.06  & 0.14 & 245.25 & 0.18 \\
VLAHFF-J071723.46+374529.8 & 109.34777 & 37.75829 & - & -          & 23.85 & 0.03   & 68.43 & 0.03 \\
VLAHFF-J071724.97+375331.3 & 109.35408 & 37.89205 & 71157 & 0.626  & 24.64  & 0.08  & 56.79 & 0.12 \\
VLAHFF-J071725.06+374714.7 & 109.35442 & 37.78742 & 54902 & 0.431  & 1.8    & 0.03 & 4.09 & 0.03 \\
VLAHFF-J071725.70+373717.3 & 109.35712 & 37.62148 & 21414 & 0.709  & 16.3   & 0.1 & 32.24 & 0.08 \\
VLAHFF-J071725.95+373352.8 & 109.35816 & 37.56468 & 18841 & 0.067  & 8.62   & 0.15 & 28.77 & 0.22 \\
VLAHFF-J071730.66+374651.1 & 109.37778 & 37.78089 & 49833 & 1.524  & 2.51   & 0.05 & 6.12 & 0.05 \\
VLAHFF-J071732.20+375619.7 & 109.38419 & 37.93882 & 86311 & 0.882  & 24.46  & 0.2 & 69.94 & 0.24 \\
VLAHFF-J071735.48+374444.8 & 109.39787 & 37.7458 & 44706 & 0.555   & 1.06   & 0.03 & 12.51 & 0.14 \\
VLAHFF-J071738.28+374650.4 & 109.40954 & 37.78067 & 50671 & 0.605  & 48.96  & 0.05  & 69.63 & 0.04 \\
VLAHFF-J071741.15+374313.7 & 109.42149 & 37.72047 & 39947 & 0.56   & 214.0  & 0.72 & 631.53 & 0.11 \\
VLAHFF-J071751.07+374440.3 & 109.46282 & 37.74454 & 44825 & 0.537  & 6.82   & 0.06 & 13.55 & 0.06 \\
VLAHFF-J071752.69+374527.0 & 109.46956 & 37.75753 & 47767 & 0.502  & 2.95   & 0.02 & 16.92 & 0.04 \\
VLAHFF-J071753.50+374209.3 & 109.47295 & 37.70261 & 36924 & 0.563  & 328.89 & 0.73 & 993.4 & 0.21 \\
VLAHFF-J071803.45+375203.4 & 109.5144 & 37.86762 & 68329 & 0.4     & 225.45 & 0.62 & 393.2 & 0.28 \\
VLAHFF-J071804.74+374852.3 & 109.51977 & 37.81453 & 56662 & 0.955  & 6.52   & 0.08 & 21.18 & 0.13 \\
VLAHFF-J071806.37+373558.1 & 109.52657 & 37.59949 & -     & -      & 141.17 & 0.44 & 130.6 & 0.20 \\
VLAHFF-J071810.75+374926.7 & 109.54479 & 37.82411 & 59007 & 0.642  & 396.47 & 0.68 & 1320.51 & 0.84 \\
VLAHFF-J071815.16+374556.3 & 109.56318 & 37.76565 & 48526 & 0.906  & 6.78   & 0.05 & 19.25 & 0.07 \\ \hline
\end{tabular}
\label{tab:extended0717}
\end{minipage}
\end{table*}

\begin{table*}
\begin{minipage}{175mm}
\centering
\caption{Positions and integrated flux densities of the extended radio sources in MACSJ1149. Please refer to the caption of Table \ref{tab:extended0416} for further details.}
\begin{tabular}{lllllllll} \hline \hline
ID & RA & DEC & CLASH ID & $z_{\mathrm{BPZ}}$ & S$_{int}^{\mathrm{HIGH}}$ & $\sigma$S$_{int}^{\mathrm{HIGH}}$& S$_{int}^{\mathrm{LOW}}$ & $\sigma$S$_{int}^{\mathrm{LOW}}$ \\
 & [deg] & [deg] & &  & [mJy] & [mJy]& [mJy]& [mJy] \\ \hline
VLAHFF-J114911.81+222049.6 & 177.29924 & 22.34713 & 47811 & 0.107 & 3.07 & 0.05 & 11.64 & 0.05 \\
VLAHFF-J114912.61+222114.4 & 177.30256 & 22.354 & 52412 & 0.175   & 3.57 & 0.06 & 19.8 & 0.08 \\
VLAHFF-J114915.00+222123.2 & 177.31251 & 22.35646 & 50105 & 0.488 & 7.77 & 0.04 & 33.26 & 0.06 \\
VLAHFF-J114917.65+221725.6 & 177.32358 & 22.29046 & 34122 & 0.708 & 27.24 & 0.03 & 63.5 & 0.08 \\
VLAHFF-J114919.43+222621.8 & 177.33099 & 22.4394 & 68009 & 0.191  & 1.48 & 0.02 & 6.672 & 0.05 \\
VLAHFF-J114922.34+222327.7 & 177.34311 & 22.39104 & 62337 & 0.24  & 1.80 & 0.03 & - & - \\
VLAHFF-J114933.09+222036.7 & 177.3879 & 22.34353 & 46902 & 0.558  & 33.74 & 0.07 & 57.73 & 0.08 \\
VLAHFF-J114933.62+221307.7 & 177.39009 & 22.21883 & -     & -     & 291.76 & 0.28 & 1316.9 & 1.15 \\
VLAHFF-J114935.51+222403.4 & 177.39796 & 22.40095 & 58850 & 0.553 & 0.49 & 0.02 & 2.09 & 0.04 \\
VLAHFF-J114936.51+222559.2 & 177.40217 & 22.43313 & 65767 & 0.739 & 5.37 & 0.02 & 74.12 & 0.05 \\
VLAHFF-J114936.83+222609.9 & 177.40349 & 22.43609 & 67239 & 0.563 & 13.49 & 0.04 & 74.21 & 0.06 \\
VLAHFF-J114939.36+222430.7 & 177.41402 & 22.40853 & 60954 & 0.561 & 17.16 & 0.02 & 69.36 & 0.03 \\
VLAHFF-J114942.53+222037.6 & 177.42725 & 22.34378 & 46364 & 0.545 & 7.77 & 0.04 & - & - \\
VLAHFF-J114952.24+222500.4 & 177.46767 & 22.4168 & 62088 & 1.123  & 1.48 & 0.02 & 6.27 & 0.03 \\
VLAHFF-J114957.22+222018.8 & 177.48843 & 22.33856 & 44764 & 0.986 & 139.16 & 0.1 & 545.25 & 0.21 \\
VLAHFF-J115003.87+221711.9 & 177.51616 & 22.28667 & 34839 & 0.229 & 1.54 & 0.07 & 5.95 & 0.1 \\
VLAHFF-J115014.53+221734.3 & 177.56056 & 22.29288 & -     & -     & 8.16 & 0.13 & 40.94 & 0.21 \\
VLAHFF-J115015.28+222052.7 & 177.56368 & 22.34797 & 48740 & 0.54  & 5.13 & 0.08 & 25.57 & 0.13 \\
VLAHFF-J115029.89+222524.5 & 177.62458 & 22.42348 & 64436 & 0.848 & 5.49 & 0.18 & 28.8 & 0.29 \\ \hline
\end{tabular}
\label{tab:extended1149}
\end{minipage}
\end{table*}

\subsection{Notes on individual extended sources}

\noindent
{\bf (1) VLAHFF-J041514.28-240934.3:} This is a low surface brightness isolated radio lobe, with a compact feature that may be a Fanaroff-Riley Type-2 \citep[FR-II;][]{fanaroff74} hotspot. It is associated with the compact core in the lower right of the panel, which also hosts a lobe to the west, however this source is not cataloged as it falls outside the primary beam cut-off. The host appears to be an elliptical.\\

\noindent
{\bf (2) VLAHFF-J041528.24-240913.7:} Double lobed structure associate with an elliptical galaxy. The host galaxy has no cataloged optical / near-infrared ID.\\

\noindent
{\bf (3) VLAHFF-J041549.94-235437.3:} Fanaroff-Riley Type-1 \citep[FR-I;][]{fanaroff74} radio source. The host is not visible in the Subaru imaging, so is possibly dust obscured.\\

\noindent
{\bf (4) VLAHFF-J041553.04-240716.5:} This is an FR-I structure, or possibly due to the low axial ratio of the lobes, {\bf a relic radio galaxy}. No compact hotspots are evident. The host is an elliptical with a redshift of 0.499. \\

\noindent
{\bf (5) VLAHFF-J041554.13-240406.9:} Two optical galaxies are enveloped by edge-brightened radio emission with a diffuse tail structure. The tabulated host is the central galaxy, at a redshift of 1.645.\\

\noindent
{\bf (6) VLAHFF-J041556.77-240434.5:} This radio source shows resolved emission apparently associated with a spiral galaxy at $z$~=~1.695. The radio emission appears to be asymmetric with respect to the disk, so it may be a jet structure associated with a rare spiral AGN, rather than star-formation driven radio emission. An alternative explanation is that the alignment with the spiral is a projection effect, and this is a radio lobe, possibly associated with the compact radio source associated with the elliptical galaxy also seen in this panel.\\

\noindent
{\bf (7) VLAHFF-J041603.56-240429.2:} A possible hybrid morphology source, the radio emission appears to be FR-I-like on the northern side, with a diffuse radio lobe on the southern side. No hotspots are visible. The optical host is a $z$~=~0.945 galaxy with unclear or disturbed optical morphology.\\

\noindent
{\bf (8) VLAHFF-J041603.63-240551.4:} This is a narrow-angle tail (NAT) radio galaxy, hosted by an elliptical galaxy at $z$~=~0.392. Tailed radio galaxies are very common in massive clusters, and the host galaxy redshift associates it with the MACSJ0416 cluster.\\

\noindent
{\bf (9) VLAHFF-J041604.63-240415.0:} A comparatively low redshift ($z$~=~0.037) elliptical galaxy, likely hosting a one-sided core-jet structure.\\

\noindent
{\bf (10) VLAHFF-J041604.84-241028.0:} Another one-sided core-jet structure, hosted by an elliptical galaxy at $z$~=~0.288. There are some artefacts associated with this source, however we believe the jet structure to be real, due to its three-contour significance, and the lack of correspondingly bright positive or negative features at the expected 120-degree positions that one would expect from a PSF-like artefact in a VLA image. \\

\noindent
{\bf (11) VLAHFF-J041605.95-235733.6:} There is an issue with the green channel in the optical imaging across this source, however it appears to be a pair of merging or conjunction of two galaxies. The radio emission exhibits a double peak and a tail. \\

\noindent
{\bf (12) VLAHFF-J041609.06-240945.0:} This another comparatively low-redshift source ($z$~=~0.025). The radio morphology exhibits two peaks, however it is most probably a core-jet structure associated with the coincident elliptical galaxy.\\

\noindent
{\bf (13) VLAHFF-J041609.16-240402.8:} A double-peaked radio source with a diffuse envelope, coincident with an elliptical galaxy with a photometric redshift of 0.438. The two radio peaks are of comparable brightness, however only one of them is coincident with an optical peak, and the detected radio structure is small compared to the size of the host. This could be a young radio jet or a double AGN.\\

\noindent
{\bf (14) VLAHFF-J041609.64-240555.3:} This source has what appears to be a one-sided core-jet structure, however it could also be a young FR-II source as the innermost radio peak is offset from the peak of the optical emission.\\

\noindent
{\bf (15) VLAHFF-J041613.94-235645.0:} This relatively low redshift ($z$~=~0.024) massive elliptical exhibits a feature in the optical imaging that resembles a spiral arm, possibly a tidal tail. The radio emission is a compact core with a diffuse envelope, possibly driven by both a central AGN and circumnuclear star formation. \\

\noindent
{\bf (16) VLAHFF-J041620.05-241008.0:} A strong compact radio source associated with a galaxy at $z$~=~1.176. The image is dynamic-range limited at this position. The linear diagonal features are likely residual sidelobes from the imperfect deconvolution, however the diffuse FR-I-like jet structure to the west may be real.\\

\noindent
{\bf (17) VLAHFF-J041621.02-235804.8:} A FR-I type galaxy, or possible wide-angle tail (WAT) source at $z$~=~1.438.\\

\noindent
{\bf (18) VLAHFF-J041625.22-240011.6:} A strong radio source associated with the core of an elliptical at $z$~=~0.516. The source exhibits a possible core-jet extension, but is dynamic range limited.\\

\noindent
{\bf (19) VLAHFF-J041628.34-235321.9:} FR-I radio morphology associated with an elliptical galaxy at $z$~=~0.358.\\

\noindent
{\bf (20) VLAHFF-J041629.22-235903.1:} Resolved radio emission associated with the disk of a spiral galaxy at $z$~=~0.574. \\

\noindent
{\bf (21) VLAHFF-J041632.03-240340.8:} A core-jet source or possibly a resolved disk at $z$~=~0.874\\

\noindent
{\bf (22) VLAHFF-J041633.46-240857.7:} One-sided core-jet structure at a redshift of 1.184.\\

\noindent
{\bf (23) VLAHFF-J041646.28-235623.4:} FR-II radio galaxy at $z$~=~3.53. \\

\noindent
{\bf (24) VLAHFF-J041648.84-240243.4:} Another bright core with a possible jet extension, but the image is dynamic range limited. The host galaxy is at a redshift of 0.831.\\

\noindent
{\bf (25) VLAHFF-J041649.20-235802.6:} A double-lobed radio galaxy at $z$~=~0.423. Host optical morphology is elliptical, with possible merging counterpart.\\

\noindent
{\bf (26) VLAHFF-J041656.65-240600.4:} Bright compact core, with possible FR-I structure but the image is dynamic range limited. The host galaxy is at $z$~=~1.148.\\

\noindent
{\bf (27) VLAHFF-J071632.28+373912.5:} Core-jet radio source associated with a compact elliptical galaxy at $z$~=~2.813. \\

\noindent
{\bf (28) VLAHFF-J071632.65+374251.6:} This is likely a core-jet source, but due the offset radio peak we assume that the true optical host is obscured by a bright foreground object at $z$~=~0.017.\\

\noindent
{\bf (29) VLAHFF-J071644.28+373956.1:} Resolved radio emission from an elliptical galaxy at $z$~=~0.076, or possibly a compact FR-I source associated with the nucleus of the host.\\

\noindent
{\bf (30) VLAHFF-J071723.46+374529.8:} This source is a NAT radio galaxy. The compact core is prominent in the C-HIGH image, however there is no compact component at that position in the S-HIGH image indicating that the core is synchrotron self-absorbed. The optical host is obscured by foreground galaxy or star in the Subaru imaging. The radio source visible to the south in the corresponding figure is unrelated.\\

\noindent
{\bf (31) VLAHFF-J071724.97+375331.3:} A highly asymmetric twin jet structure, or possible a WAT with a resolved out or otherwise non-detected jet. The optical data places the host at $z$~=~0.626.\\

\noindent
{\bf (32) VLAHFF-J071725.06+374714.7:} Resolved radio emission associated with a spiral galaxy at $z$~=~0.431. The radio morphology shows a possible spiral arm structure.\\

\noindent
{\bf (33) VLAHFF-J071725.70+373717.3:} Diffuse double-lobed radio source at $z$~=~0.709, with no clear sign of the hotspot emission at the lobe heads.\\

\noindent
{\bf (34) VLAHFF-J071725.95+373352.8:} An intriguing source at redshift 0.067, showing compact jets associated with either a large edge-on spiral, or an elliptical with a prominent dust lane. \\

\noindent
{\bf (35) VLAHFF-J071730.66+374651.1:} This source is likely a compact FR-II radio galaxy at $z$~=~1.524.\\

\noindent
{\bf (36) VLAHFF-J071732.20+375619.7:} A tailed radio galaxy, or extremely one-sided source. At a redshift of 0.882 it is not a NAT or WAT associated with the foreground cluster.\\

\noindent
{\bf (37) VLAHFF-J071735.48+374444.8:} A compact core-jet source, seen to be embedded in the bright halo emission of the MACS0717 cluster. With a photometric redshift of 0.555 the host galaxy is likely to be a cluster member.\\

\noindent
{\bf (38) VLAHFF-J071738.28+374650.4:} A FR-I source at $z$~=~0.605. The northern jet has a significant extension that changes direction with respect to the jet emission close to the core, however there is no southern counterpart to this very extended feature.\\

\noindent
{\bf (39) VLAHFF-J071741.15+374313.7:} This source is a NAT with twisted jets that are clearly visible in the C-HIGH image. The host appears to be an elliptical galaxy with a photometric redshift of 0.56. \\

\noindent
{\bf (40) VLAHFF-J071751.07+374440.3:} A WAT source at $z$~=~0.537, likely associated with a cluster member.\\

\noindent
{\bf (41) VLAHFF-J071752.69+374527.0:} This is a compact FR-I source at $z$~=~0.502.\\

\noindent
{\bf (42) VLAHFF-J071753.50+374209.3:} A double-lobed source with a pair of diffuse inner hotspots, possibly a restarted AGN. The host is at $z$~=~0.563.\\

\noindent
{\bf (43) VLAHFF-J071803.45+375203.4:} A head-tail radio galaxy at $z$~=~0.4 with very diffuse, low axial ratio jet emission. Could possibly be a NAT or WAT seen in projection. \\

\noindent
{\bf (44) VLAHFF-J071804.74+374852.3:} Possible hybrid morphology (FR-I/FR-II) radio source at $z$~=~0.955.\\

\noindent
{\bf (45) VLAHFF-J071806.37+373558.1:} A bent tail (NAT/WAT) radio source with a clear optical host in the Subaru images but no cataloged counterpart. \\

\noindent
{\bf (46) VLAHFF-J071810.75+374926.7:} This source is possibly a one-sided FR-I with a large opening angle, but the image is dynamic range limited due to the bright radio core. The host galaxy is at $z$~=~0.642.\\

\noindent
{\bf (47) VLAHFF-J071815.16+374556.3:} The radio emission shows either a core-jet source, or possibly the resolved disk of the optical host galaxy at $z$~=~0.906. \\

\noindent
{\bf (48) VLAHFF-J114911.81+222049.6:} This radio source is likely driven by circumnuclear star formation in an elliptical galaxy at $z$~=~0.107. \\

\noindent
{\bf (49) VLAHFF-J114912.61+222114.4:} Resolved radio emission associated with the disk of a spiral galaxy at $z$~=~0.175.\\

\noindent
{\bf (50) VLAHFF-J114915.00+222123.2:} FR-I structure associated with a redshift 0.488 elliptical galaxy.\\

\noindent
{\bf (51) VLAHFF-J114917.65+221725.6:} Double compact radio sources, both within the redshift 0.708 host. Possible double nucleus.\\

\noindent
{\bf (52) VLAHFF-J114919.43+222621.8:} The morphology of this source suggests it is resolved star-formation driven radio emission at $z$~=~0.191.\\

\noindent
{\bf (53) VLAHFF-J114922.34+222327.7:} This is a bright FR-I radio source on the periphery of the MACS1149 field, however with a photometric redshift of 0.24 it is likely not associated with the cluster itself. The host appears to be a large, low surface brightness elliptical galaxy. We discuss this source further in Section \ref{sec:relics}.\\

\noindent
{\bf (54) VLAHFF-J114933.09+222036.7:} This source is a WAT with a redshift of 0.558, likely associated with the MACS1149 cluster.\\

\noindent
{\bf (55) VLAHFF-J114933.62+221307.7:} There is no cataloged counterpart for the host galaxy of this source, however a strong candidate is visible in the Subaru imaging. The radio morphology is a disturbed, asymmetric FR-II structure with prominent radio plumes.\\

\noindent
{\bf (56) VLAHFF-J114935.51+222403.4:} This object is a one-sided core-jet source associated with a redshift 0.553 galaxy, likely a cluster member.\\

\noindent
{\bf (57) VLAHFF-J114936.51+222559.2:} A compact FR-II source with a $z$~=~0.739 host galaxy.\\

\noindent
{\bf (58) VLAHFF-J114936.83+222609.9:} A tailed radio galaxy, likely a NAT or WAT seen in projection. The optical counterpart is clear, with a cataloged redshift of 0.563, likely associating it with the MACS1149 cluster.\\

\noindent 
{\bf (59) VLAHFF-J114939.36+222430.7:} This appears to be a bent FR-I source at $z$~=~0.536.\\

\noindent
{\bf (60) VLAHFF-J114942.53+222037.6:} Another NAT at $z$~=~0.545. This source appears to be embedded in the eastern radio relic associated with the MACS1149 cluster. See Section \ref{sec:relics} for more details.\\

\noindent
{\bf (61) VLAHFF-J114952.24+222500.4:} Three components are visible in this compact source, which is possibly a young FR-II with a redshift of 1.123. \\

\noindent
{\bf (62) VLAHFF-J114957.22+222018.8:} The radio emission from this strong source appears to be resolved in multiple directions, but the image is dynamic range limited at this point, so this must be interpreted with care. The host galaxy is at a redshift of 0.986.\\

\noindent
{\bf (63) VLAHFF-J115003.87+221711.9:} This is diffuse radio emission at low signal to noise ratio. Likely driven by star formation, it is associated with the core of a spiral galaxy at $z$~=~0.229.\\

\noindent
{\bf (64) VLAHFF-J115014.53+221734.3:} No optical counterpart is visible for this source. Given the morphology it is likely to be an isolated radio lobe. The morphology also suggests the most likely core counterpart is the compact source VLAHFF-J115008.52+221733.8 ($z$~=~0.25, and not visible in this panel). At this redshift the projected separation is 328 kpc, which is not unreasonable for an FR-II. No western counterpart lobe is visible on the other side of the putative core.\\

\noindent
{\bf (65) VLAHFF-J115015.28+222052.7:} This is a FR-I source associated with an elliptical galaxy at $z$~=~0.54, likely associated with MACSJ1149.\\

\noindent
{\bf (66) VLAHFF-J115029.89+222524.5:} The radio emission is aligned perpendicular to the major axis of the elliptical host galaxy ($z$~=~0.848), and is likely a compact pair of radio jets.\\

\begin{figure*}[ht!]
\centering
\includegraphics[width= \textwidth]{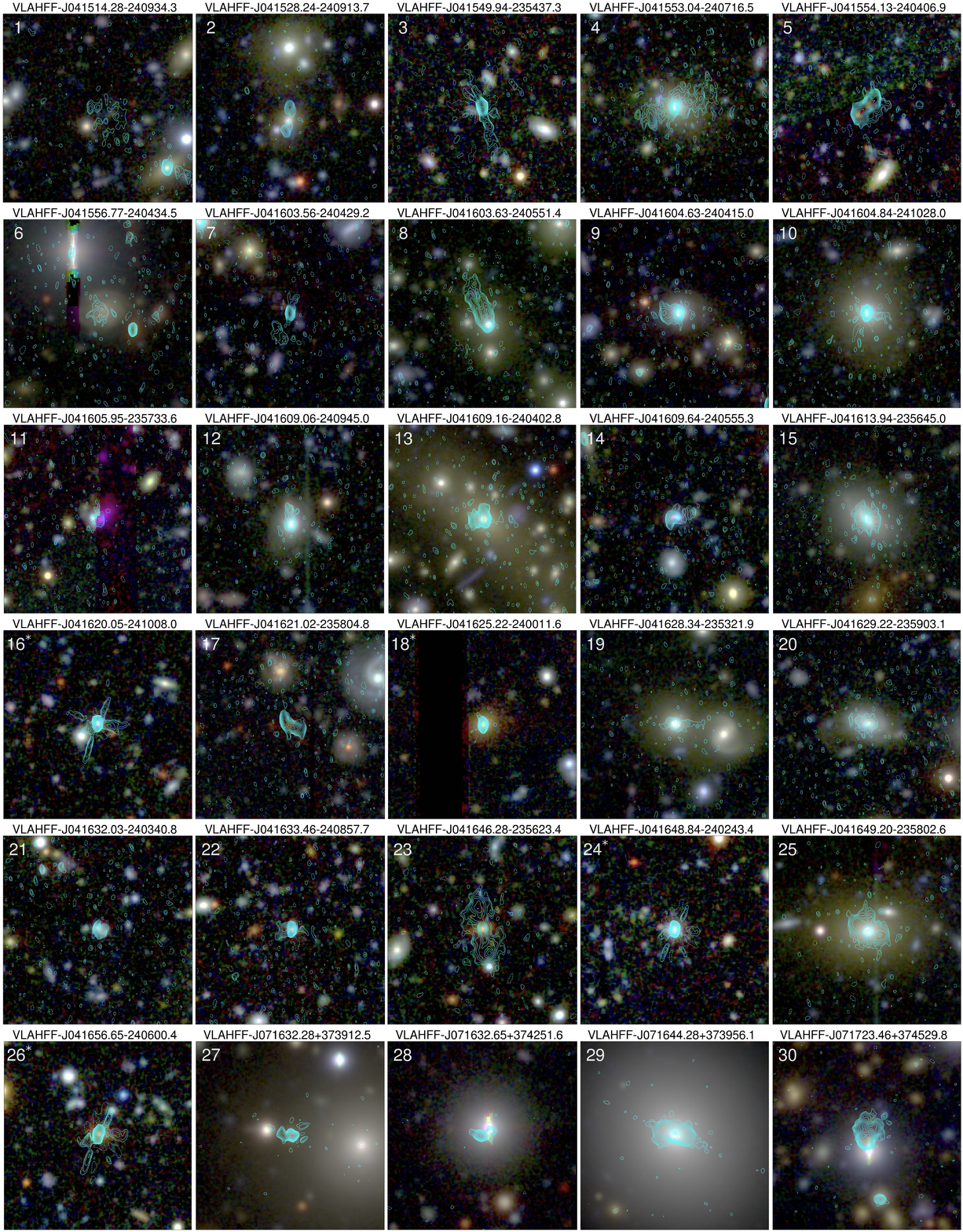}
\caption{Sources with complex morphologies. Please refer to final panel for the full caption.
\label{fig:extended1}}
\end{figure*}

\setcounter{figure}{12}
\begin{figure*}[ht!]
\centering
\includegraphics[width= \textwidth]{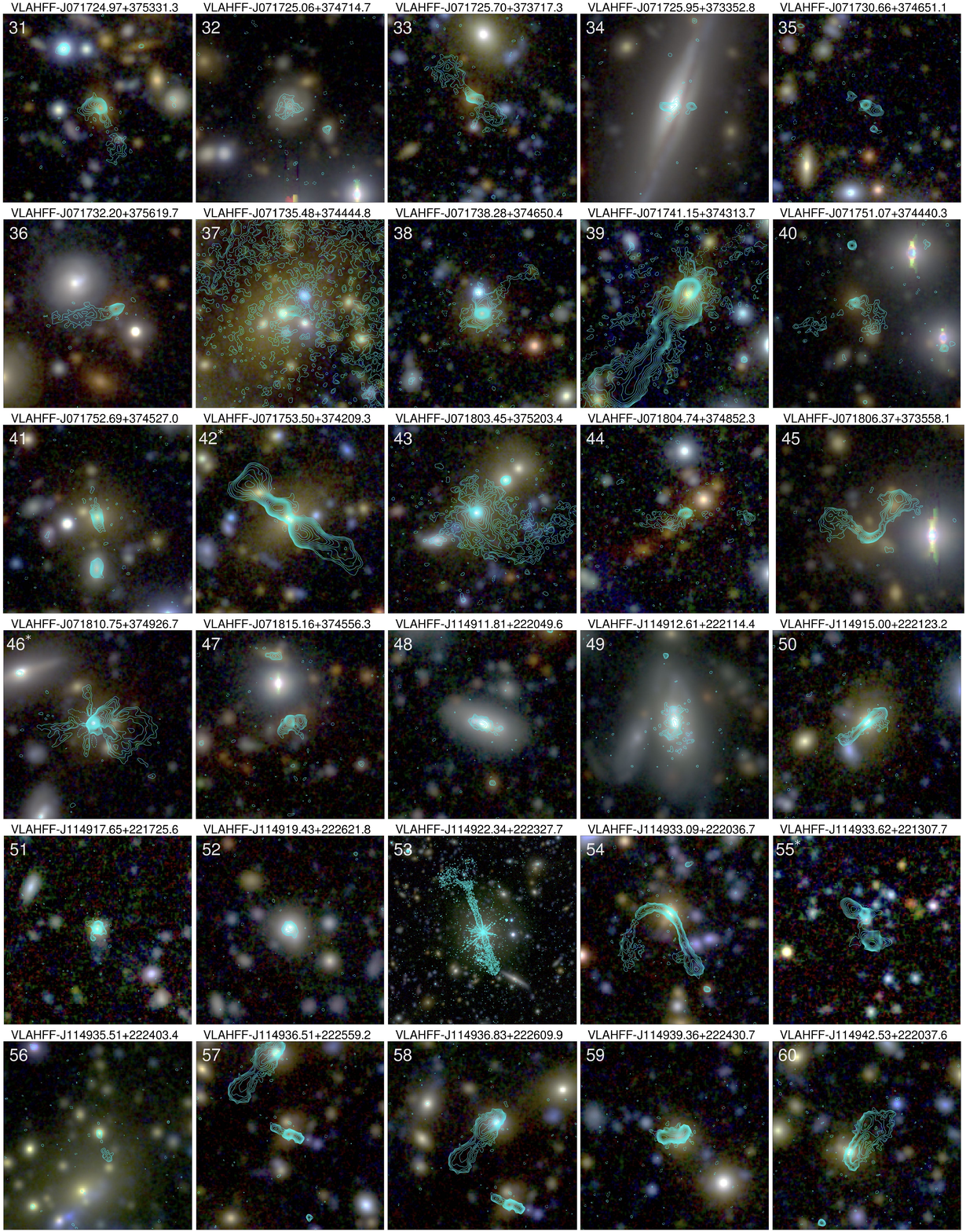}
\caption{\emph{continued}.
\label{fig:extended2}}
\end{figure*}

\setcounter{figure}{12}
\begin{figure}[h]
\centering
\includegraphics[width=\columnwidth]{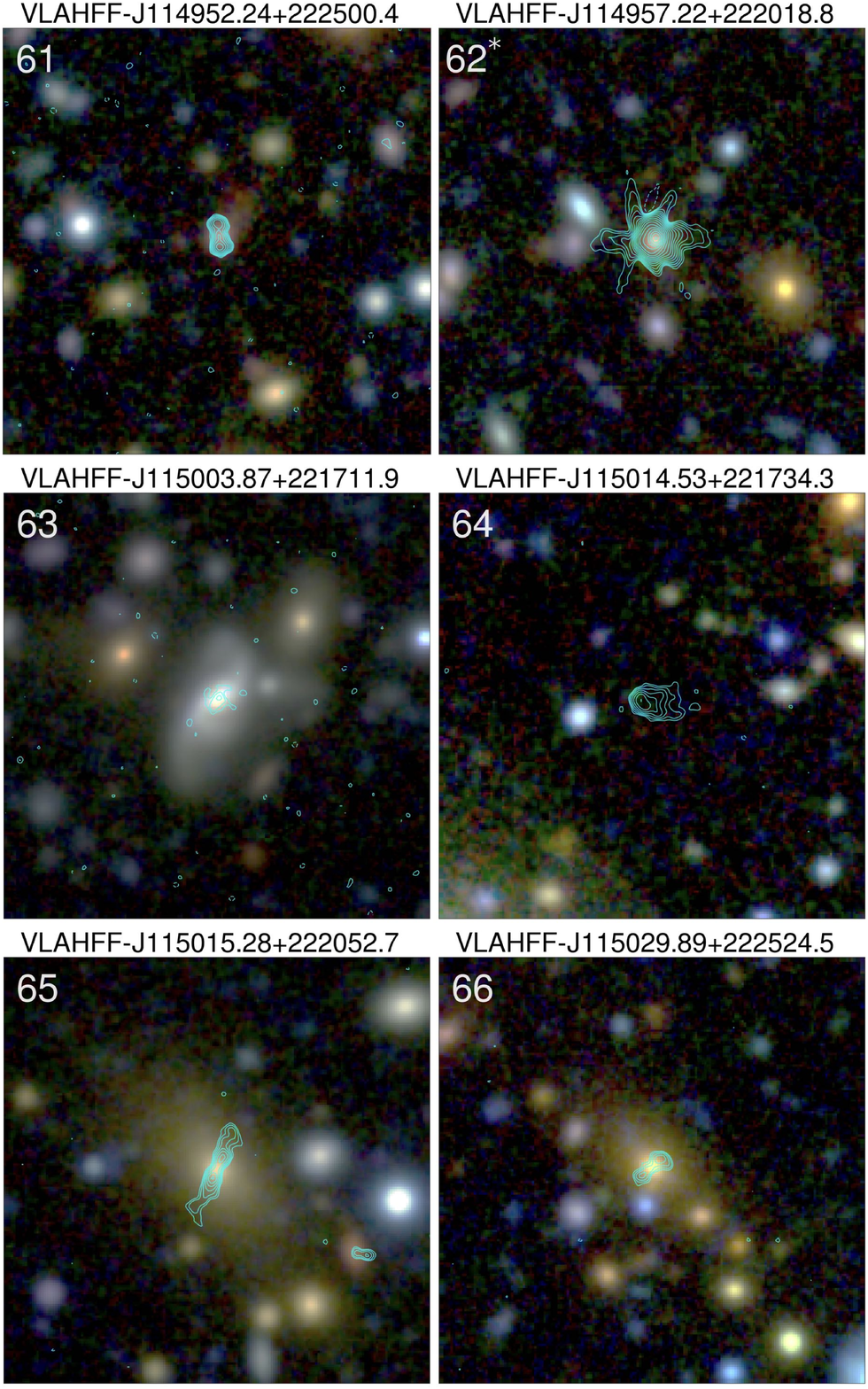}
\caption{Sources with complex morphologies. Contour {\bf are from the S-HIGH image}, with levels of 1.7 $\times$ (1, $\sqrt{2}$, 2, 2$\sqrt{2}$, 4, 4$\sqrt{2}$, 8, 8$\sqrt{2}$, \ldots) $\mu$Jy beam$^{-1}$, and with a single (dashed line) negative contour at $-$1.7 $\mu$Jy beam$^{-1}$, except for the brightest sources for which the base level is increased from $\pm$1.7 $\mu$Jy beam$^{-1}$ to $\pm$6 $\mu$Jy beam$^{-1}$. Panels with these increased contour values are marked with an asterisk next to the panel number. The extent of each panel is 28$''$.8, with the exception of VLAHFF-J114922.34+222327.7 (panel 53) in Figure \ref{fig:extended2}, which has been expanded to cover 108$''$ to show the full extent of the source. Note that these images have not been primary beam corrected in order to maintain the uniform noise and contouring scheme. Primary beam corrected flux density measurements are available in Tables \ref{tab:extended0416}, \ref{tab:extended0717}, and \ref{tab:extended1149}. Note also that the brightness-weighted centroid is given as the source position by ProFound, and so the images may not necessarily be centered on the host galaxy.}
\label{fig:extended3}
\end{figure}


\begin{thebibliography}{}


\bibitem[\protect\citeauthoryear{Astropy Collaboration et al.}{2013}]{astropy:2013} Astropy Collaboration, Robitaille T.~P., Tollerud E.~J., Greenfield P., Droettboom M., Bray E., Aldcroft T., et al., 2013, A\&A, 558, A33. doi:10.1051/0004-6361/201322068

\bibitem[\protect\citeauthoryear{Astropy Collaboration et al.}{2018}]{astropy:2018} Astropy Collaboration, Price-Whelan A.~M., Sip{\H{o}}cz B.~M., G{\"u}nther H.~M., Lim P.~L., Crawford S.~M., Conseil S., et al., 2018, AJ, 156, 123. doi:10.3847/1538-3881/aabc4f


\bibitem[\protect\citeauthoryear{Bonafede, et al.}{2012}]{bonafede12} Bonafede A., et al., 2012, MNRAS, 426, 40

\bibitem[\protect\citeauthoryear{Bonafede, et al.}{2017}]{bonafede17} Bonafede A., et al., 2017, MNRAS, 470, 3465

\bibitem[\protect\citeauthoryear{Bonafede, et al.}{2018}]{bonafede18} Bonafede A., et al., 2018, MNRAS, 478, 2927

\bibitem[\protect\citeauthoryear{Brammer, van Dokkum, \& Coppi}{2008}]{brammer08} Brammer G.~B., van Dokkum P.~G., Coppi P., 2008, ApJ, 686, 1503. doi:10.1086/591786

\bibitem[\protect\citeauthoryear{Briggs}{1995}]{briggs95} Briggs D.~S., 1995, AAS

\bibitem[\protect\citeauthoryear{Caminha, et al.}{2017}]{caminha17} Caminha G.~B., et al., 2017, A\&A, 600, A90

\bibitem[Castellano et al.(2016)]{castellano16} Castellano, M., Amor{\'{\i}}n, R., Merlin, E., et al.\ 2016, \aap, 590, A31

\bibitem[Condon(1997)]{condon97} Condon, J.~J.\ 1997, \pasp, 109, 166 

\bibitem[\protect\citeauthoryear{Delvecchio, et al.}{2017}]{delvecchio17} Delvecchio I., et al., 2017, A\&A, 602, A3

\bibitem[Di Criscienzo et al.(2017)]{dicriscienzo17} Di Criscienzo, M., Merlin, E., Castellano, M., et al.\ 2017, \aap, 607, A30

\bibitem[\protect\citeauthoryear{Diego, et al.}{2015}]{diego15} Diego J.~M., Broadhurst T., Zitrin A., Lam D., Lim J., Ford H.~C., Zheng W., 2015, MNRAS, 451, 3920

\bibitem[\protect\citeauthoryear{Dudzevi{\v{c}}i{\={u}}t{\.{e}} et al.}{2020}]{dudzeviciute20} Dudzevi{\v{c}}i{\={u}}t{\.{e}} U., Smail I., Swinbank A.~M., Stach S.~M., Almaini O., da Cunha E., An F.~X., et al., 2020, MNRAS, 494, 3828. doi:10.1093/mnras/staa769

\bibitem[\protect\citeauthoryear{Ebeling, Edge \& Henry}{2001}]{ebeling01} Ebeling H., Edge A.~C., Henry J.~P., 2001, ApJ, 553, 668

\bibitem[Ebeling et al.(2014)]{ebeling14} Ebeling, H., Ma, C.-J., \& Barrett, E.\ 2014, \apjs, 211, 21 

\bibitem[\protect\citeauthoryear{Fanaroff \& Riley}{1974}]{fanaroff74} Fanaroff B.~L., Riley J.~M., 1974, MNRAS, 167, 31P

\bibitem[\protect\citeauthoryear{Gim, et al.}{2019}]{gim19} Gim H.~B., et al., 2019, ApJ, 875, 80

\bibitem[\protect\citeauthoryear{Hale et al.}{2019}]{hale19} Hale C.~L., Robotham A.~S.~G.,Davies L.~J.~M., Jarvis M.~J., Driver S.~P., Heywood I., 2019, MNRAS, 487, 3971

\bibitem[\protect\citeauthoryear{Heywood, et al.}{2016}]{heywood16} Heywood I., et al., 2016, MNRAS, 460, 4433

\bibitem[\protect\citeauthoryear{Heywood et al.}{2020}]{heywood20} Heywood I., Hale C.~L., Jarvis M.~J., Makhathini S., Peters J.~A., Sebokolodi M.~L.~L., Smirnov O.~M., 2020, MNRAS, 496, 3469. doi:10.1093/mnras/staa1770

\bibitem[\protect\citeauthoryear{Hunter}{2007}]{hunter07} Hunter J.~D., 2007, CSE, 9, 90. doi:10.1109/MCSE.2007.55

\bibitem[\protect\citeauthoryear{Huynh, et al.}{2015}]{huynh15} Huynh M.~T., Bell M.~E., Hopkins A.~M., Norris R.~P., Seymour N., 2015, MNRAS, 454, 952

\bibitem[\protect\citeauthoryear{Huynh, et al.}{2020}]{huynh20} Huynh M.~T., Seymour N., Norris R.~P., Galvin T., 2020, MNRAS, 491, 3395

\bibitem[\protect\citeauthoryear{Ivison et al.}{2007}]{ivison07} Ivison R.~J., Greve T.~R., Dunlop J.~S., Peacock J.~A., Egami E., Smail I., Ibar E., et al., 2007, MNRAS, 380, 199. doi:10.1111/j.1365-2966.2007.12044.x

\bibitem[\protect\citeauthoryear{Jackson}{2011}]{jackson11} Jackson N., 2011, ApJL, 739, L28. doi:10.1088/2041-8205/739/1/L28

\bibitem[\protect\citeauthoryear{Jauzac, et al.}{2016}]{jauzac16} Jauzac M., et al., 2016, MNRAS, 457, 2029

\bibitem[\protect\citeauthoryear{Kriek et al.}{2018}]{kriek18} Kriek M., van Dokkum P.~G., Labb{\'e} I., Franx M., Illingworth G.~D., Marchesini D., Quadri R.~F., et al., 2018, ascl.soft. ascl:1803.008

\bibitem[Lotz et al.(2017)]{lotz17} Lotz, J.~M., Koekemoer, A., Coe, D., et al.\ 2017, \apj, 837, 97 

\bibitem[\protect\citeauthoryear{Kawamata, et al.}{2016}]{kawamata16} Kawamata R., Oguri M., Ishigaki M., Shimasaku K., Ouchi M., 2016, ApJ, 819, 114

\bibitem[\protect\citeauthoryear{Limousin, et al.}{2016}]{limousin16} Limousin M., et al., 2016, A\&A, 588, A99

\bibitem[Medezinski et al.(2013)]{medezinski13} Medezinski, E., Umetsu, K., Nonino, M., et al.\ 2013, \apj, 777, 43 

\bibitem[\protect\citeauthoryear{Magnelli, et al.}{2013}]{magnelli13} Magnelli B., et al., 2013, A\&A, 553, A132

\bibitem[\protect\citeauthoryear{Merlin, et al.}{2015}]{merlin15} Merlin E., et al., 2015, A\&A, 582, A15

\bibitem[\protect\citeauthoryear{Merlin, et al.}{2016}]{merlin16} Merlin E., et al., 2016, A\&A, 590, A30

\bibitem[\protect\citeauthoryear{McKinnon, et al.}{2019}]{mckinnon19} McKinnon M., Beasley A., Murphy E., Selina R., Farnsworth R., Walter A., 2019, BAAS, 51, 81

\bibitem[McMullin et al.(2007)]{mcmullin07} McMullin, J.~P., Waters, B., Schiebel, D., Young, W., \& Golap, K.\ 2007, Astronomical Data Analysis Software and Systems XVI, 376, 127 

\bibitem[Mohan \& Rafferty(2015)]{mohan15} Mohan, N., \& Rafferty, D.\ 2015, Astrophysics Source Code Library, ascl:1502.007 

\bibitem[\protect\citeauthoryear{Murphy, et al.}{2011}]{murphy11} Murphy E.~J., Chary R.-R., Dickinson M., Pope A., Frayer D.~T., Lin L., 2011, ApJ, 732, 126

\bibitem[Murphy et al.(2017)]{murphy17} Murphy, E.~J., Momjian, E., Condon, J.~J., et al.\ 2017, \apj, 839, 35 

\bibitem[\protect\citeauthoryear{Murphy, et al.}{2018}]{murphy18} Murphy E.~J., et al., 2018, ASPC, 517, 3, ASPC..517

\bibitem[Offringa et al.(2014)]{offringa14} Offringa, A.~R., McKinley, B., Hurley-Walker, N., et al.\ 2014, \mnras, 444, 606 

\bibitem[\protect\citeauthoryear{Ogrean et al.}{2015}]{ogrean15} Ogrean G.~A., van Weeren R.~J., Jones C., Clarke T.~E., Sayers J., Mroczkowski T., Nulsen P.~E.~J., et al., 2015, ApJ, 812, 153. doi:10.1088/0004-637X/812/2/153


\bibitem[Perley(2016)]{perley16}Perley, R., 2016, VLA Expansion Project Memoranda, 195 (available here: \href{https://library.nrao.edu/public/memos/evla/EVLAM_195.pdf}{https://library.nrao.edu/public/memos/evla/EVLAM\_195.pdf})

\bibitem[\protect\citeauthoryear{Rajpurohit et al.}{2020}]{rajpurohit20} Rajpurohit K., Wittor D., van Weeren R.~J., Vazza F., Hoeft M., Rudnick L., Locatelli N., et al., 2020, arXiv, N


\bibitem[Robotham et al.(2018)]{robotham18} Robotham, A.~S.~G., Davies, L.~J.~M., Driver, S.~P., et al.\ 2018, \mnras, 476, 3137 

\bibitem[\protect\citeauthoryear{Schmidt, et al.}{2014}]{schmidt14} Schmidt K.~B., et al., 2014, ApJL, 782, L36

\bibitem[\protect\citeauthoryear{Shipley et al.}{2018}]{shipley18} Shipley H.~V., Lange-Vagle D., Marchesini D., Brammer G.~B., Ferrarese L., Stefanon M., Kado-Fong E., et al., 2018, ApJS, 235, 14. doi:10.3847/1538-4365/aaacce

\bibitem[\protect\citeauthoryear{Singh et al.}{2014}]{singh14} Singh V., Beelen A., Wadadekar Y., Sirothia S., Ishwara-Chandra C.~H., Basu A., Omont A., et al., 2014, A\&A, 569, A52. doi:10.1051/0004-6361/201423644

\bibitem[\protect\citeauthoryear{Smol{\v{c}}i{\'c}, et al.}{2017}]{smolcic17} Smol{\v{c}}i{\'c} V., et al., 2017, A\&A, 602, A1

\bibitem[Umetsu et al.(2014)]{umetsu14} Umetsu, K., Medezinski, E., Nonino, M., et al.\ 2014, \apj, 795, 163 

\bibitem[\protect\citeauthoryear{van Weeren et al.}{2009}]{vanweeren09} van Weeren R.~J., R{\"o}ttgering H.~J.~A., Br{\"u}ggen M., Cohen A., 2009, A\&A, 505, 991. doi:10.1051/0004-6361/200912528

\bibitem[\protect\citeauthoryear{van Weeren, et al.}{2016}]{vanweeren16} van Weeren R.~J., et al., 2016, \apj, 817, 98

\bibitem[\protect\citeauthoryear{van Weeren, et al.}{2017}]{vanweeren17} van Weeren R.~J., et al., 2017, ApJ, 835, 197

\bibitem[\protect\citeauthoryear{Vanzella, et al.}{2014}]{vanzella14} Vanzella E., et al., 2014, ApJL, 783, L12

\bibitem[\protect\citeauthoryear{Wang, et al.}{2019}]{wang19} Wang T., et al., 2019, Natur, 572, 211




\end{thebibliography}
\end{document}